\documentclass[
reprint,
superscriptaddress,
amsmath,amssymb,
aps,
prl,
twocolumn
]{revtex4-2}

\usepackage{graphicx}
\usepackage{dcolumn}
\usepackage{bm}
\usepackage[breaklinks]{hyperref}
\hypersetup{colorlinks=true, linkcolor=blue, citecolor=blue, filecolor=blue, urlcolor=blue}
\graphicspath{{figures}}
\usepackage{bm}
  
\usepackage{graphicx,subfigure}
\usepackage{epsfig}
\usepackage{bm}
\usepackage{dcolumn}
\usepackage{color}
\usepackage{physics}
\usepackage{float}
\makeatletter
\newcommand*{\rom}[1]{\expandafter\@slowromancap\romannumeral #1@}
\makeatother
\usepackage{lipsum}
\usepackage{subfigure}
\usepackage{dsfont}
\usepackage{enumitem}
\usepackage{tikz}
\usepackage[normalem]{ulem}

\begin{document}
\title{Exploring Quantum Phase Transitions Beyond Landau Theory on a Quantum Processor}

\author{Ruizhe Shen}
\email{e0554228@u.nus.edu}
\affiliation{Department of Physics, National University of Singapore, Singapore 117551}
\author{Yin Zhong}
\email{zhongy@lzu.edu.cn}
\affiliation{School of Physical Science and Technology $\&$ Key Laboratory for Magnetism and Magnetic Materials of the MoE, Lanzhou University, Lanzhou 730000, China} %
\affiliation{Lanzhou Center for Theoretical Physics, Key Laboratory of Theoretical Physics of Gansu Province, Lanzhou 730000, China}
\author{Ching Hua Lee}
\email{phylch@nus.edu.sg}
\affiliation{Department of Physics, National University of Singapore, Singapore 117551}
\date{\today}

\begin{abstract}
Direct continuous quantum phase transitions between phases that break distinct symmetries challenge the conventional Landau–Ginzburg–Wilson paradigm and provide a framework for deconfined quantum criticality. Here, we investigate a Landau-forbidden ferromagnetic-to-valence-bond-solid (FM–VBS) transition in a frustrated one-dimensional spin chain using superconducting quantum processors. Combining matrix-product-state-assisted circuit preparation with quantum-hardware measurements, we resolve the suppression of long-range ferromagnetic correlations and the redistribution of competing-order fluctuations across the transition. Distance-resolved correlations and finite-size analyses reveal signatures consistent with critical behavior. We further extend our investigation to nonequilibrium dynamics, directly preparing dimerized states and tracking the evolution of VBS order and joint FM–VBS fluctuations under local unitary interactions. Our results establish a versatile quantum-hardware framework for probing competing orders, critical correlations, and many-body dynamics beyond conventional Landau theory.
\end{abstract}

 \maketitle

{\bf Introduction.}~
Understanding quantum phase transitions beyond the conventional
Landau--Ginzburg--Wilson paradigm is a central problem in quantum
many-body physics
\cite{Sachdev2011QPT,Cardy1996Scaling,Cui2025CPLReview,
Sreejith2019SO5,Lee2019ShastrySutherland,Sandvik2020Consistent,
Zhao2020Multicritical,Kuklov2008FirstOrder,Melko2008Scaling,
Lou2009AFMVBS,Sandvik2010LogCorrections,Kaul2012LargeN,
Jiang2008FirstOrder,Shen2023YangLeePRL}.
Within the Landau framework, phases are classified by local order
parameters and their associated symmetry-breaking patterns, while
continuous transitions are described by the long-wavelength
fluctuations of these order parameters.
A continuous transition is therefore naturally associated with
the continuous disappearance of an order parameter.
However, when two phases break unrelated symmetries, their
order parameters are independent within the conventional Landau
description.
A direct continuous transition between them would require the
disappearance of one order and the onset of the other at the
same critical point, which is generally not expected without
additional fine-tuning.
Instead, the transition is typically first order, or the two
ordered phases are separated by an intermediate phase, potentially
one in which both orders coexist
\cite{Senthil2004Science,Senthil2004PRB}.

Deconfined quantum criticality provides a prominent route beyond this
expectation, in which the critical theory can involve fractionalized degrees
of freedom, topological defects, emergent gauge fields, and dual descriptions
beyond those of the conventional order parameters
\cite{Senthil2004Science,Senthil2004PRB,Senthil2005JPSJ,Wang2017PRX,Qin2017PRX}.
The canonical example is the transition between N\'eel antiferromagnetic and
valence-bond-solid orders, where numerical and field-theoretic studies have
found unconventional scaling, multiple characteristic length scales, and
possible enlarged emergent symmetries
\cite{Sandvik2007PRL,Nahum2015PRX,Nahum2015PRL,Shao2016Science,
Wang2017PRX,Ma2018PRB,Ma2019PRL}.
The precise nature of the putative DQCP nevertheless remains under active
debate, with recent studies employing disorder operators, entanglement
diagnostics, conformal bootstrap, fuzzy-sphere regularization, and
order-parameter coupling to distinguish genuine continuous criticality from
pseudocritical or weakly first-order behavior
\cite{Wang2022SciPost,Zhao2022PRL,DEmidio2023SciPost,Zhou2024PRX,
ChesterSu2024PRL,Deng2024PRL,DEmidio2024PRL,Hawashin2025NatCommun,
Shu2025NatCommun}.
Recent experimental and quantum-simulation developments have further
broadened this landscape, including pressure- and field-tuned
Shastry--Sutherland materials, extended Shastry--Sutherland models with
emergent symmetry, and Rydberg-array proposals for realizing and probing
Landau-forbidden criticality
\cite{Cui2023Science,Liu2024PRL,Lee2023PRL,Guo2025CommPhys,
Bombieri2025PRL,Cui2025CPLReview}.

Most studies of deconfined quantum criticality have focused on two-dimensional
spin systems, particularly transitions between antiferromagnetic and
valence-bond-solid (VBS) orders
\cite{Senthil2004Science,Senthil2004PRB,Sandvik2007PRL,Nahum2015PRL,
Wang2017PRX,Qin2017PRX,Senthil2023Review,Cui2025CPLReview}.
One-dimensional analogues offer a complementary setting for studying
Landau-forbidden criticality with greater numerical control and are naturally
connected to programmable quantum-simulation platforms
\cite{Roberts2019PRB,Huang2019PRB,ZhangLevin2023PRL,Romen2024SciPostCore,
Lee2023PRL,Huang2020KibbleZurekDQCP,Yang2023EmergentSymmetry,
Pronk2025ArXiv}.
Recent quantum-hardware studies of critical and strongly correlated
one-dimensional states further motivate this direction
\cite{Dupont2022QuantumCriticality,Miao2026EntanglementScaling,
Koyluoglu2026CentralCharge,Smith2022TopologicalTransition,
Pennington2026SPT100,shen2025robust,ChenShen2023AKLTSciPost,Shen2025FermiSkinNatCommun,ChenShen2026DTCQST,Shen2026ASPTArXiv,Shen2024ManyBodyScarsPRL}.
In one dimension, frustrated and symmetry-enriched spin chains can host
direct continuous transitions between phases with distinct broken
symmetries, providing controlled realizations of Landau-forbidden critical
behavior
\cite{Mudry2019PRB,JiangMotrunich2019PRB,Roberts2019PRB,Huang2019PRB,
XiYu2022CPB,XiYu2022JPhysA,ZhangLevin2023PRL,Romen2024SciPostCore,
Luo2023KitaevDQCP,Yang2023EmergentSymmetry,Chen2025NoninvertibleDQCP,
Pronk2025ArXiv}.
A particularly relevant example is the transition between a ferromagnetic
(FM) phase and a VBS phase: the FM phase breaks an internal Ising spin-flip
symmetry, whereas the VBS phase breaks one-site lattice translation symmetry
\cite{JiangMotrunich2019PRB,Roberts2019PRB,Huang2019PRB,XiYu2022CPB}.
A direct FM--VBS transition therefore provides a one-dimensional realization
of deconfined criticality, with descriptions involving dual variables,
self-duality, emergent continuous symmetry, and continuously varying
critical exponents
\cite{JiangMotrunich2019PRB,Roberts2019PRB,Huang2019PRB,
Roberts2021PRB,XiYu2022JPhysA,Yang2023PRB}.

Our results establish a quantum-hardware approach to probing
a Landau-forbidden FM--VBS transition and the dynamics of
competing orders in one-dimensional spin chains.
Using tensor-network-assisted circuit preparation, we
experimentally resolve the suppression of long-range
ferromagnetic correlations and the redistribution of
competing-order fluctuations across the transition.
Distance-resolved correlations and finite-size analyses
further reveal signatures consistent with the critical
behavior predicted by DMRG.
Beyond the ground-state transition, we implement
Floquet-type driven dynamics on chains of up to 96 qubits,
observing the suppression of initial bond dimerization
accompanied by enhanced ferromagnetic fluctuations.
These results demonstrate that present-day
quantum processors can access both equilibrium signatures
of unconventional quantum criticality and nonequilibrium
dynamics of competing orders, providing a versatile
platform for investigating strongly correlated many-body
systems and phenomena beyond the conventional Landau
paradigm
\cite{Senthil2004Science,Senthil2004PRB,
Senthil2023Review,Lee2023PRL,Bombieri2025PRL}.

\begin{figure}
 \centering
 \includegraphics[width=0.99\linewidth]{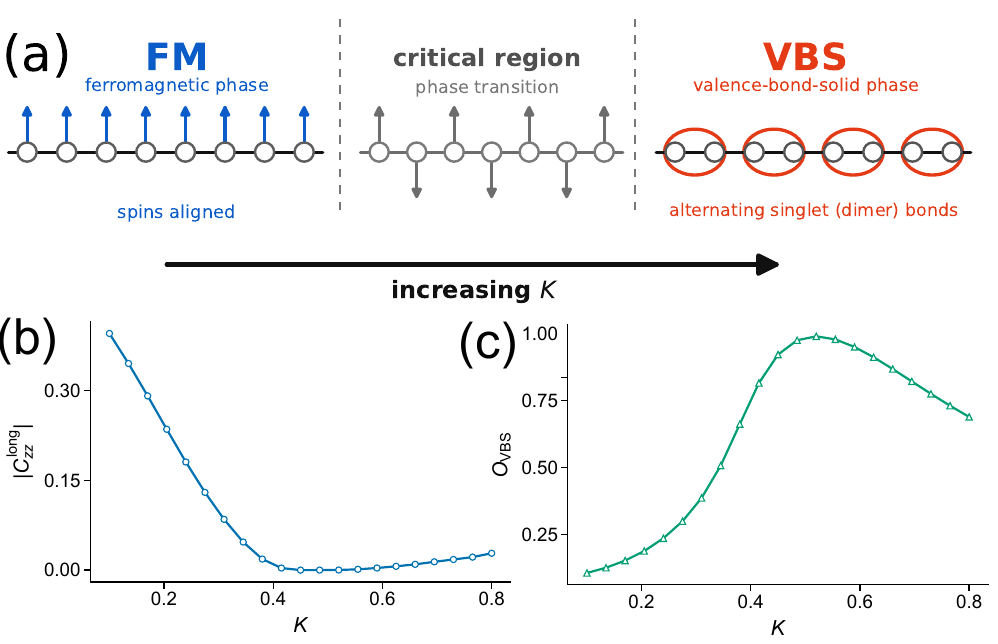}
\caption{
\textbf{Ferromagnetic-to-VBS transition analogous to a deconfined phase transition.}
\textbf{(a)} Schematic illustration of the frustration-driven quantum phase transition obtained by increasing $K$ for $J_z>J_x$. At small $K$, the system is in a ferromagnetic (FM) phase characterized by long-range spin alignment. At large $K$, frustration favors alternating strong bonds, giving rise to a valence-bond-solid (VBS) phase with spontaneously broken lattice translation symmetry.
\textbf{(b)} Absolute value of the long-distance ferromagnetic correlator $|C_{zz}^{\mathrm{long}}|$ [Eq.~\ref{eq:Czz_long_def}] as a function of $K$. Its rapid suppression near $K_c$ signals the destruction of ferromagnetic order.
\textbf{(c)} VBS order parameter $O_{\mathrm{VBS}}$ [Eq.~\ref{eq:ovbs_def}] as a function of $K$. Its growth in the same critical region signals the onset of bond dimerization and the VBS phase. Panels \textbf{(b)} and \textbf{(c)} show DMRG results for $L=32$, $J_z=1$, and $J_x=0.8$.
}
\label{fig:FM_VBS_transition}
 \label{fig:fig1}
\end{figure}

{\bf Models.~} We investigate the one-dimensional frustrated spin-$1/2$ XZ chain \cite{JiangMotrunich2019PRB,Roberts2019PRB,Huang2019PRB},
\begin{equation}
H\!\!=\!\!\!
\sum_{j}
\left[
- J_x \sigma^x_j \sigma^x_{j+1}
- J_z \sigma^z_j \sigma^z_{j+1}
\!+\! K_{2x} \sigma^x_j \sigma^x_{j+2}
\!+\! K_{2z} \sigma^z_j \sigma^z_{j+2}
\right],
\label{eq:JM_model}
\end{equation}
where $\sigma_j^{x,z}$ are Pauli operators.
The nearest-neighbor couplings $J_x$ and $J_z$ favor ferromagnetic alignment, while the next-nearest-neighbor couplings $K_{2x}$ and $K_{2z}$ generate competing antiferromagnetic interactions.
Unless otherwise specified, we restrict our analysis to the symmetric frustration line $K_{2x}=K_{2z}\equiv K$, so that the frustration is controlled by the single parameter $K$.

At small $K$, the nearest-neighbor interactions stabilize a spin-ordered FM phase. Increasing $K$ strengthens the competing next-nearest-neighbor interactions, suppressing ferromagnetic order and eventually favoring a dimerized VBS phase with alternating strong and weak bonds \cite{MajumdarGhosh1969JMP,JiangMotrunich2019PRB}.
These phases break distinct symmetries: the FM phase breaks an internal Ising spin-flip symmetry, whereas the VBS phase breaks one-site lattice translation symmetry \cite{JiangMotrunich2019PRB,Roberts2019PRB,Huang2019PRB}.
A direct continuous transition between such unrelated ordered phases is not generic within the conventional Landau framework and can therefore be viewed as a one-dimensional analogue of a deconfined phase transition \cite{Senthil2004Science,Senthil2004PRB,JiangMotrunich2019PRB,Roberts2019PRB,Huang2019PRB,Mudry2019PRB,Roberts2021PRB}.
The corresponding DMRG signatures are summarized in Fig.~\ref{fig:fig1} \cite{White1992PRL,Schollwock2011AnnPhys,JiangMotrunich2019PRB}.
Figure~\ref{fig:fig1}(a) schematically shows the evolution from the FM state at weak frustration to the dimerized VBS state at strong frustration.

We diagnose the FM--VBS transition using complementary DMRG order parameters, as shown in Figs.~\ref{fig:fig1}(b,c).
For the ferromagnetic order, we fix a reference site at the chain center and define the distance-resolved correlator
$C_{zz}(r)=\langle Z_{i_0}Z_{i_0+r}\rangle$, with $i_0=L/2$.
The long-distance component is obtained by averaging over separations larger than $L/4$,
\begin{equation}
C_{zz}^{\mathrm{long}}
=
\frac{1}{N_r}
\sum_{r>L/4}
\left\langle
Z_{i_0} Z_{i_0+r}
\right\rangle ,
\label{eq:Czz_long_def}
\end{equation}
where $N_r$ is the number of separations included in the average.
A finite $|C_{zz}^{\mathrm{long}}|$ indicates long-range ferromagnetic order, while its suppression signals the loss of the FM phase.
For the competing VBS order, we define the local bond expectation value
$B_j=-J_x\langle X_jX_{j+1}\rangle-J_z\langle Z_jZ_{j+1}\rangle$ and extract its staggered component
\begin{equation}
O_{\mathrm{VBS}}
=
\frac{1}{N_{\mathrm{b}}}
\sum_{j\in \mathcal{B}_{\mathrm{bulk}}}
(-1)^j B_j ,
\label{eq:ovbs_def}
\end{equation}
where $\mathcal{B}_{\mathrm{bulk}}$ denotes the set of bulk bonds and $N_{\mathrm{b}}$ is the number of such bonds.
A finite $|O_{\mathrm{VBS}}|$ signals alternating strong and weak bonds, corresponding to spontaneous dimerization.
Fig.~\ref{fig:fig1}(b) shows that $|C_{zz}^{\mathrm{long}}|$ remains finite in the weak-frustration regime but is rapidly suppressed near the critical region around $K_c\approx0.4$, while Fig.~\ref{fig:fig1}(c) shows that $|O_{\mathrm{VBS}}|$ grows over the same region in which the long-range ferromagnetic correlations are suppressed. The smooth evolution of both order parameters reflects finite-size rounding of the underlying continuous transition. The simultaneous loss of FM order and the onset of VBS order therefore identify the finite-size critical region associated with the direct continuous FM--VBS transition \cite{JiangMotrunich2019PRB,Roberts2019PRB,Huang2019PRB}.

\begin{figure}
 \centering
 \includegraphics[width=0.99\linewidth]{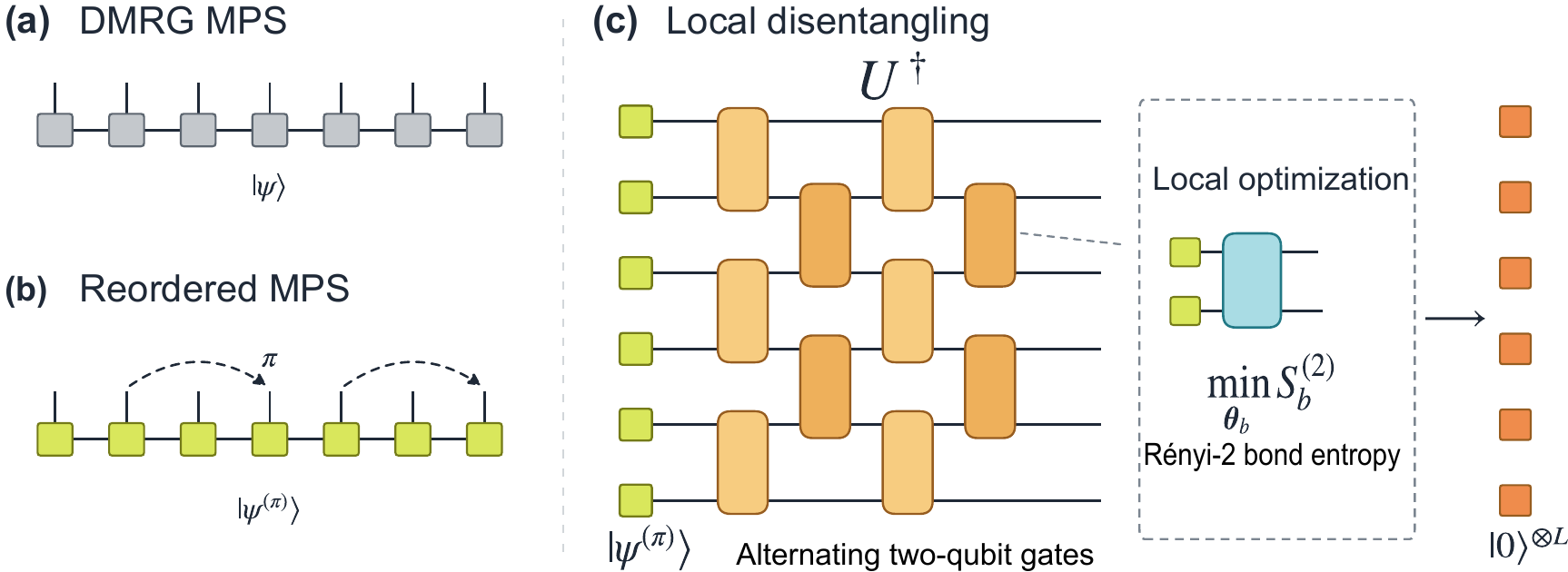}

\caption{
\textbf{Circuit compilation for quantum-hardware simulation.}
Schematic of the tensor-network-assisted state-preparation procedure, which exploits the correlation and entanglement structure of many-body states to construct shallow quantum circuits.
\textbf{(a)} The DMRG ground state $|\psi\rangle$ with open boundary conditions is represented as an MPS in the original physical-site order.
\textbf{(b)} A site permutation $\pi$ is introduced to bring strongly correlated sites closer together. The MPS tensors are rearranged through adjacent site swaps, yielding the reordered target state $|\psi^{(\pi)}\rangle$.
\textbf{(c)} A parameterized disentangling circuit $U^\dagger$ is constructed to approximately map the reordered MPS to the product state $|0\rangle^{\otimes L}$ [Eq.~\ref{disent}]. As illustrated in the inset, the two-qubit blocks are optimized by locally minimizing the entanglement entropy across the corresponding MPS bonds [Eq.~\ref{eq:local_bond_cost}]. The inverse circuit $U$ is then used to prepare the target state on quantum hardware.
Further details are provided in the Supplementary Material.
}

 \label{fig:fig2}
\end{figure}

{\bf Quantum circuit implementation.}~ To access the FM--VBS transition on quantum hardware, we approximate each DMRG ground state using a parameterized quantum circuit.
For each value of $K$, the ground state obtained with bond dimension $\chi=128$ is represented in canonical MPS form and used as the target for circuit optimization \cite{White1992PRL,Schollwock2011AnnPhys,Hastings2007JSTAT}.
This tensor-network-assisted strategy maps the MPS target to a shallow variational circuit compatible with the hardware connectivity \cite{Ran2020PRA,Rudolph2024QST,Malz2024PRL,Smith2024PRXQuantum}.
We optimize against a reordered MPS target $|\psi(\pi)\rangle$, where circuit qubit $q_a$ represents the original physical site $\pi_a$.
The permutation brings strongly correlated physical sites closer in the circuit ordering, allowing a nearest-neighbor ansatz to capture the dominant correlations at reduced circuit depth \cite{Preskill2018Quantum,Kandala2017Nature,Cerezo2021NRP}.
In practice, $\pi$ is selected from a small set of correlation-guided candidate orderings, avoiding an exhaustive search over all permutations and introducing only polynomial optimization overhead.

The circuit-construction workflow is summarized in Fig.~\ref{fig:fig2}.
Starting from the DMRG ground state $|\psi\rangle$ in the natural lattice ordering [Fig.~\ref{fig:fig2}(a)], we introduce a permutation $\pi=(\pi_0,\pi_1,\ldots,\pi_{L-1})$, where $\pi_a$ denotes the original physical site placed at circuit position $a$ [Fig.~\ref{fig:fig2}(b)].
The ordering is chosen to reduce the circuit distance between strongly correlated sites.
We then optimize a disentangling circuit satisfying
\begin{equation}
\label{disent}
U^\dagger(\boldsymbol{\theta})|\psi(\pi)\rangle
\simeq
|0\rangle^{\otimes L},
\end{equation}
so that the corresponding hardware preparation circuit generates
$U(\boldsymbol{\theta})|0\rangle^{\otimes L}$.
As illustrated in Fig.~\ref{fig:fig2}(c), the circuit is organized in alternating layers of nonoverlapping two-qubit blocks. 
For a bond $b$, we optimize the corresponding two-qubit
block $U_b(\boldsymbol{\theta}_b)$ by minimizing the
second R\'enyi entanglement entropy across that bond.
The local cost function is defined as
\begin{equation}
\mathcal{C}_b(\boldsymbol{\theta}_b)
=
S_b^{(2)}\!\left[
U_b(\boldsymbol{\theta}_b)|\psi_b\rangle
\right]
=
-\ln\left[
\sum_{\alpha}
\lambda_{b,\alpha}^{4}(\boldsymbol{\theta}_b)
\right],
\label{eq:local_bond_cost}
\end{equation}
where $|\psi_b\rangle$ denotes the current MPS before
applying the block, and
$\lambda_{b,\alpha}(\boldsymbol{\theta}_b)$ are the
normalized Schmidt coefficients across bond $b$
after its application.
The block parameters are determined by
$\boldsymbol{\theta}_b^\star
=
\operatorname*{arg\,min}_{\boldsymbol{\theta}_b}
\mathcal{C}_b(\boldsymbol{\theta}_b)$.
Each optimized block is applied to the MPS before
proceeding to the next bond, progressively reducing
the local entanglement.
This construction replaces a global optimization
over all circuit parameters with a sequence of local
entanglement-minimization steps.
The resulting two-qubit blocks are arranged in
alternating layers of nonoverlapping gates, allowing
parallel execution within each layer on quantum hardware.
Further details are provided in the Supplementary Material. The DMRG target lies in the even symmetry sectors of
$G_x=\prod_{j=1}^{L}\sigma_j^x$ and
$G_z=\prod_{j=1}^{L}\sigma_j^z$, with
$G_x=+1$ and $G_z=+1$.
Because $G_z$ is diagonal in the computational basis,
we postselect the measured $Z$-basis bitstrings onto
the $G_z=+1$ sector before extracting the order
parameters.

\begin{figure}
 \centering
 \includegraphics[width=0.99\linewidth]{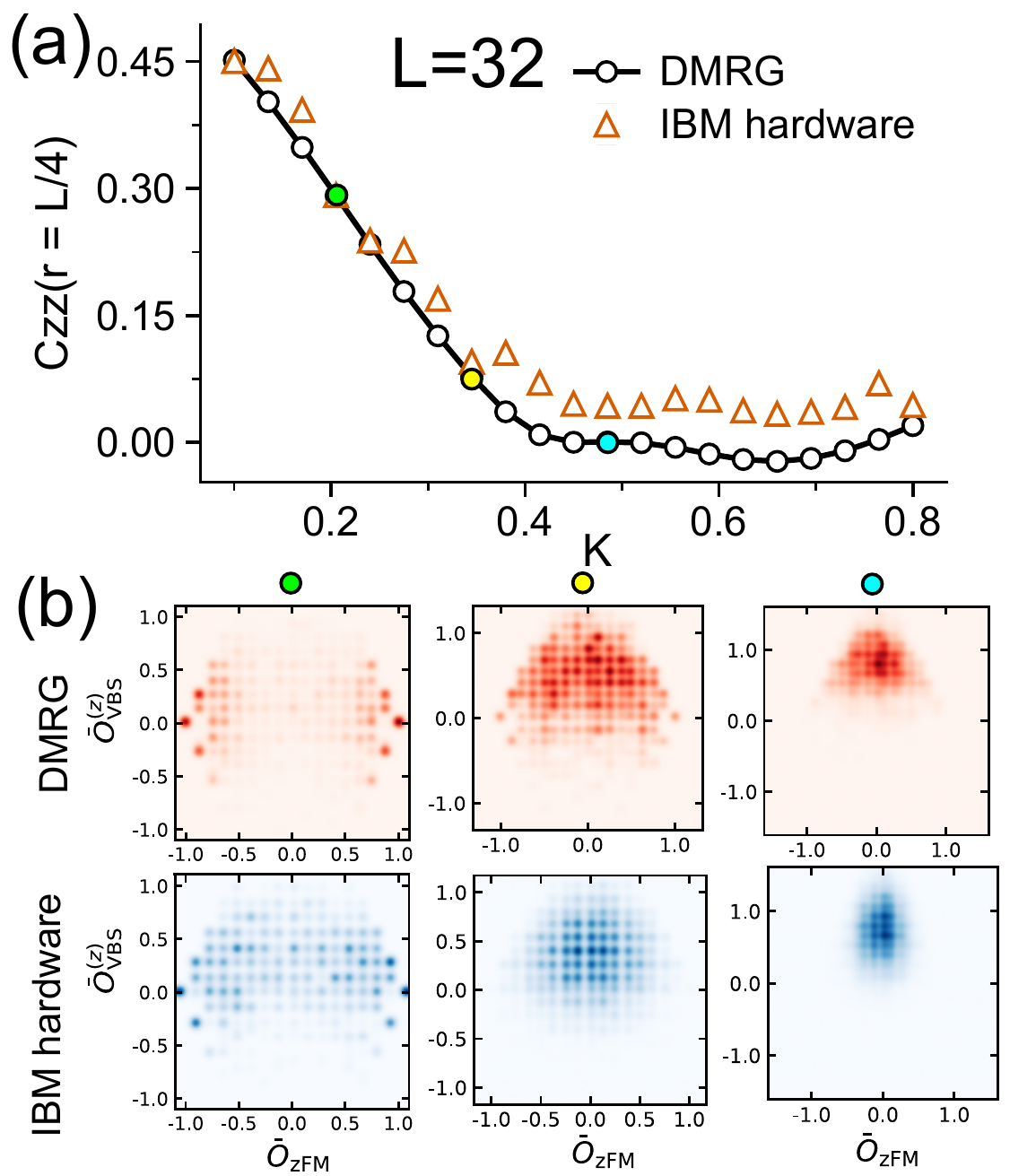}
 \caption{
\textbf{Quantum hardware measurement of the ferromagnetic correlator across the FM--VBS transition.}
\textbf{(a)} Bulk spin correlator $C_{zz}(i=L/2,j=L/2+L/4)=\langle\sigma^z_{L/2}\sigma^z_{3L/4}\rangle$ as a function of the frustration strength $K$. Black circles denote DMRG results, and orange triangles denote IBM quantum-hardware measurements. The correlator remains finite in the small-$K$ ferromagnetic regime and is rapidly suppressed as $K$ approaches the transition region near $K_c\approx0.4$, signaling the loss of long-range ferromagnetic order.
\textbf{(b)} Joint probability distributions of the competing order parameters $(\bar O_{\rm zFM},\bar O'_{\rm VBS})$ obtained from DMRG and IBM Quantum data at the marked values of $K$. In the FM regime, the weight is concentrated mainly along the $\bar O_{\rm zFM}$ direction. Near the critical region, it develops an arc-like profile for $\widetilde{\boldsymbol O}
\propto
R(\cos\theta,\sin\theta)$, while on the VBS side the weight shifts toward the VBS direction. Further details are provided in the Supplementary Material. We set $J_z=1$ and $J_x=0.8$ throughout.
}
 \label{fig:fig3}
\end{figure}

{\bf {Quantum hardware} measurement of the FM--VBS transition.~} 
Although the corresponding observables can be evaluated directly from the MPS,
the central challenge for quantum simulation is to realize such complex
many-body states on hardware with experimentally accessible circuit depth.
Our recompilation procedure addresses this preparation bottleneck by encoding
the target states into shallow circuits while preserving their relevant
many-body correlations. We next implement these circuits on IBM quantum
hardware and probe the ferromagnetic correlations across the transition.
The measured observable is the bulk correlator
\begin{equation}
C_{zz}(r=L/4)
=
\left\langle
\sigma^z_{L/2}
\sigma^z_{L/2+L/4}
\right\rangle,
\label{eq:Czz_L4_hardware}
\end{equation}
which probes spin correlations over a distance proportional to the system size.
This correlator is finite in the ferromagnetic phase and is suppressed when long-range FM order is destroyed.

Fig.~\ref{fig:fig3}(a) compares the IBM hardware measurements with the corresponding DMRG results across the scan of $K$.
At weak frustration, the measured correlator remains positive and finite, consistent with $z$-ferromagnetic order.
As $K$ increases, $C_{zz}(r=L/4)$ decreases rapidly and approaches zero near the critical region, $K_c\approx0.4$.
The measurements therefore resolve the loss of long-range ferromagnetic correlations across the FM--VBS transition.
The largest deviations occur near the transition and in the large-$K$ regime, where the correlator is already small and is therefore more sensitive to finite circuit depth, sampling fluctuations, and hardware noise.

To further characterize the FM--VBS transition, we examine the joint
probability distribution of the competing order parameters obtained from
$Z$-basis snapshots
\cite{JiangMotrunich2019PRB,Huang2019PRB,Lee2023PRL}.
For IBM hardware, each snapshot corresponds to a single experimental shot,
whereas for DMRG the snapshots are sampled from the $Z$-basis Born
distribution of the corresponding MPS. For each sampled spin configuration
$\mathbf{s}$, we evaluate the spatially averaged ferromagnetic estimator
$\bar O_{\rm zFM}(\mathbf{s})$ and the snapshot-compatible dimer estimator
$\bar O'_{\rm VBS}(\mathbf{s})$. The resulting ensemble of pairs
$\bigl(\bar O_{\rm zFM},\bar O'_{\rm VBS}\bigr)$ defines the joint distribution
$P(\bar O_{\rm zFM},\bar O'_{\rm VBS})$. The dimer estimator transforms in the
same way as the VBS order under the microscopic symmetries and therefore has
a nonzero overlap with the leading VBS scaling field
\cite{Lee2023PRL}.

For a periodic chain, the Hamiltonian is invariant under one-site translation
$\mathsf{T}_x$ and the on-site $Z_2^x\times Z_2^z$ spin rotations:
\begin{align}
g_x:\quad
&
\bar O_{\rm zFM}\rightarrow-\bar O_{\rm zFM},
&
\bar O'_{\rm VBS}\rightarrow\bar O'_{\rm VBS},
\nonumber\\
\mathsf{T}_x:\quad
&
\bar O_{\rm zFM}\rightarrow\bar O_{\rm zFM},
&
\bar O'_{\rm VBS}\rightarrow-\bar O'_{\rm VBS}.
\label{eq:fm_vbs_symmetry_transformations}
\end{align}
Thus, the FM and VBS orders break distinct microscopic $Z_2$ symmetries and
are not related by an exact continuous symmetry of the lattice Hamiltonian.
Accordingly, in the FM phase the joint distribution is concentrated near
$(\bar O_{\rm zFM},\bar O'_{\rm VBS})\simeq(\pm m,0)$ when both
spin-flip-related sectors are sampled, whereas in a translation-invariant
periodic system the two VBS sectors would appear near $(0,\pm d)$.
For the open chains considered here, one-site translation is not an exact
symmetry and the boundaries generally select one of the two dimerization
patterns. The measured distribution therefore need not be symmetric under
$\bar O'_{\rm VBS}\rightarrow-\bar O'_{\rm VBS}$.

Near the FM--VBS critical point, the FM and VBS fluctuations are expected to form the two components of an emergent $O(2)$ order parameter. Because the corresponding lattice estimators generally have different nonuniversal scales, we examine their rescaled joint distribution, $\widetilde{\boldsymbol O}=(\widetilde O_{\rm FM},\widetilde O_{\rm VBS})$. Each measurement snapshot yields a different realization of this vector, with a fluctuating magnitude $R=|\widetilde{\boldsymbol O}|$, giving rise to a two-dimensional distribution rather than a fixed-radius arc.

For the finite open chains considered here, boundary-induced VBS pinning favors one sign of $\widetilde O_{\rm VBS}$, while the two Ising-related FM sectors remain accessible. As the system approaches the critical region, the distribution evolves from two FM-dominated peaks into a broad structure extending toward the positive VBS direction [Fig.~\ref{fig:fig3}(b)]. The DMRG results exhibit a weak arc-like tendency with substantial radial fluctuations, whereas the IBM measurements show a more diffuse distribution. This behavior is qualitatively consistent with the intertwining of FM and VBS fluctuations expected near the deconfined critical point, although the joint distribution alone does not establish criticality or emergent $O(2)$ symmetry \cite{JiangMotrunich2019PRB,Huang2019PRB,Lee2023PRL}.

\begin{figure}
 \centering
 \includegraphics[width=0.99\linewidth]{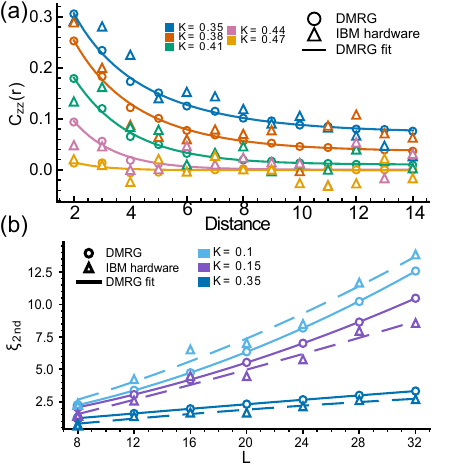}
\caption{
\textbf{Measurements of distance-resolved correlations and a finite-size correlation length on quantum hardware.}
\textbf{(a)} Ferromagnetic correlator
$C_{zz}(r)=\langle Z_{L/2}Z_{L/2+r}\rangle$
for $L=32$ at selected $K$.
Open circles and triangles show DMRG and IBM hardware results, respectively.
Solid curves are fits to the DMRG data.
\textbf{(b)} Second-moment correlation length
$\xi_{\mathrm{2nd}}$ [Eq.~\ref{eq:xi2nd_main}] versus system size $L$ for selected $K$,
with the same marker and line conventions.  At $K=0.35$, $\xi_{\mathrm{2nd}}$ increases approximately linearly with
$L$, $\xi_{\mathrm{2nd}}\propto L$.
Here, $J_z=1.0$ and $J_x=0.8$.
}
 \label{fig.fig4}
\end{figure}

{\bf Finite-size correlations.~}  Near a continuous quantum critical point, the ferromagnetic correlations are expected to exhibit algebraic decay in the thermodynamic limit. In finite chains, however, their spatial dependence is modified by the system size and boundary conditions, as well as by finite circuit depth in the hardware implementation and finite bond dimension in the DMRG calculations \cite{Cardy1996Scaling,Sachdev2011QPT,Tagliacozzo2008PRB,Pollmann2009PRL,Schollwock2011AnnPhys}. We examine the distance dependence of the ferromagnetic correlator $C_{zz}(r)=\langle Z_{L/2}Z_{L/2+r}\rangle$, as shown in Fig.~\ref{fig.fig4}(a) for $L=32$. At smaller $K$, the correlations remain appreciable over long distances, reflecting robust ferromagnetic correlations. As $K$ increases, their magnitude is progressively suppressed, consistent with the loss of ferromagnetic order toward the VBS regime. The IBM hardware measurements reproduce the same systematic dependence on $K$ as the DMRG results.

To further characterize the finite-size correlation structure without relying on this multi-parameter real-space fit, we evaluate a second-moment correlation-length estimator from the static structure factor \cite{PelissettoVicari2002},
\begin{equation}
\xi_{\mathrm{2nd}}(L)
=
\frac{1}{2\sin(\pi/L)}
\left[
\frac{S_{zz}(0)}{S_{zz}(2\pi/L)}-1
\right]^{1/2},
\label{eq:xi2nd_main}
\end{equation}
where $S_{zz}(q)=L^{-1}\sum_{i,j}e^{iq(i-j)}\langle Z_iZ_j\rangle$. Here, $\xi_{\mathrm{2nd}}$ is determined by the relative weights of the zero- and finite-momentum correlations, in contrast to $\xi_{\mathrm{fit}}$, which is extracted from the real-space decay. The two quantities characterize the same underlying correlation structure but are not expected to coincide quantitatively in finite systems. For the open chains considered here, we use $\xi_{\mathrm{2nd}}$ as a finite-size structure-factor estimator.

As shown in Fig.~\ref{fig.fig4}(b), $\xi_{\mathrm{2nd}}$ increases with system size over the investigated range. At $K=0.35$, its growth is approximately linear, $\xi_{\mathrm{2nd}}(L)\propto L$, indicating that the low-momentum correlation structure exhibits a pronounced system-size dependence. Such behavior is consistent with finite-size scaling near a continuous critical point, although the accessible sizes do not allow this scaling alone to establish criticality. The IBM hardware measurements capture the same qualitative size dependence as the DMRG calculations, demonstrating that these finite-size correlation signatures remain experimentally accessible despite hardware noise. Further details of the analysis and error mitigation are provided in the Supplementary Material.

\begin{figure*}[t]
    \centering
    \includegraphics[width=0.9\linewidth]
    {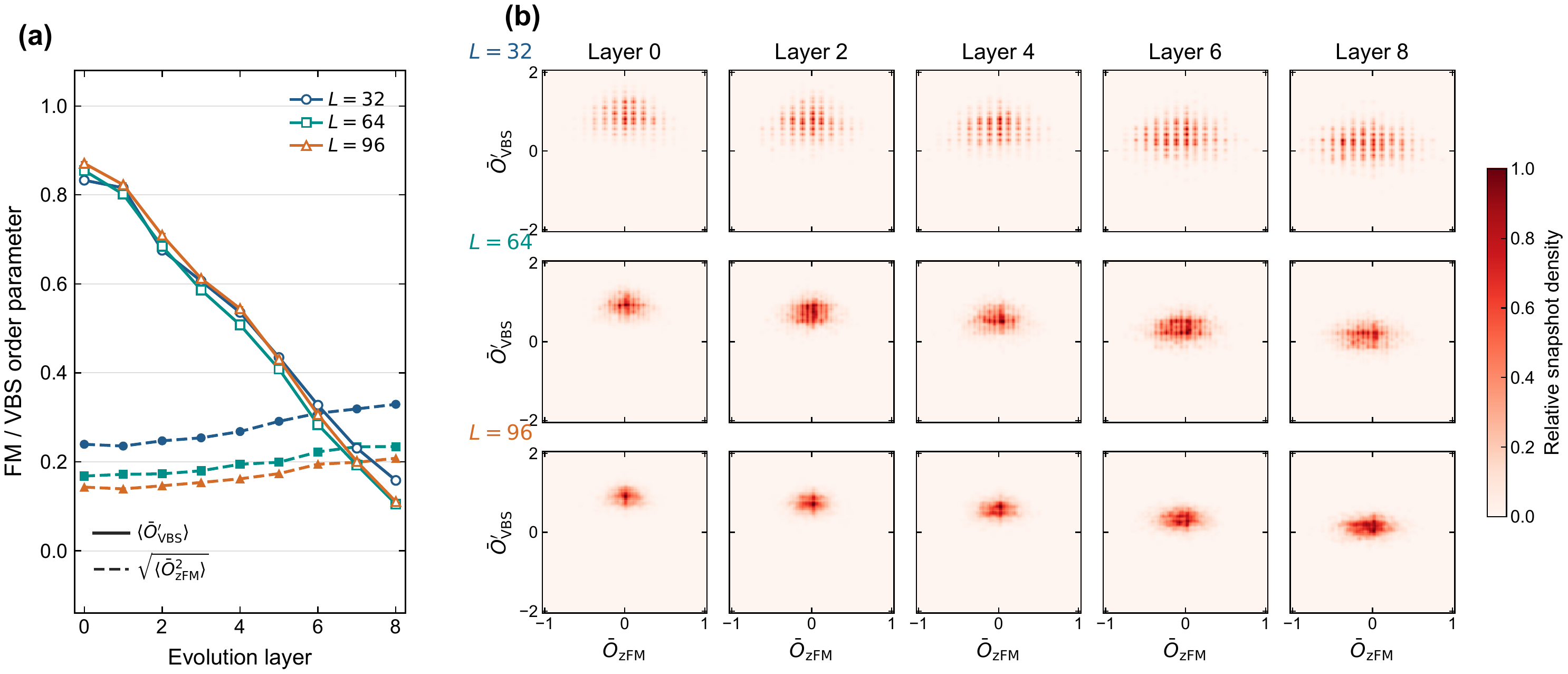}
    \caption{
    \textbf{Floquet-engineered evolution of competing
    FM and VBS correlations on quantum hardware.}
    \textbf{(a)} Evolution of the mean VBS snapshot estimator
    $\langle\bar O'_{\mathrm{VBS}}\rangle$ (solid lines)
    and FM fluctuation amplitude
    $\sqrt{\langle\bar O_{\mathrm{zFM}}^2\rangle}$
    (dashed lines) for $L=32$, $64$, and $96$.
    The initial state is a product of Bell-pair dimers,
    $|\psi_0\rangle=\bigotimes_{j=0}^{L/2-1}
    (|00\rangle+|11\rangle)_{2j,2j+1}/\sqrt{2}$.
    The dynamics follow the layer-dependent Floquet
    circuit defined in Eq.~\eqref{eq:vbs_floquet},
    with $J_d=J_z=1$, $J_x=0.5$,
    $T=4$, and eight driving cycles.
    \textbf{(b)} Joint snapshot distributions
    $P(\bar O_{\mathrm{zFM}},\bar O'_{\mathrm{VBS}})$
    at depths $d=0,2,4,6$, and $8$ (left to right)
    for $L=32$, $64$, and $96$ (top to bottom).
    The suppression of VBS order accompanied by
    enhanced FM fluctuations demonstrates that
    the driven dynamics destabilize the initial
    dimerized correlations.}
    \label{fig:vbs_dynamics}
\end{figure*}

{\bf Floquet-engineered dynamics of competing FM and VBS orders.~}
To investigate the dynamical competition between FM and VBS
correlations, we initialize a spin-$1/2$ chain in an idealized
dimerized state,
$|\psi_0\rangle=\bigotimes_{j=0}^{L/2-1}
|\Phi^+\rangle_{2j,2j+1}$,
where $|\Phi^+\rangle=(|00\rangle+|11\rangle)/\sqrt{2}$.
The initial state exhibits perfect intradimer $ZZ$
correlations but no correlations between different dimers.

We implement a layer-dependent Floquet-type drive
consisting of alternating even- and odd-bond interactions.
The unitary for the $d$th driving cycle is
\begin{equation}
U_F(s_d)=U_{\mathrm{o}}(s_d)U_{\mathrm{e}}(s_d),
\quad
U_p(s)=\prod_{j\in\mathcal{B}_p}
e^{-i\tau h_{j,j+1}^{(p)}(s)},
\label{eq:vbs_floquet}
\end{equation}
where $p=\mathrm{e},\mathrm{o}$ denotes the even- and
odd-bond sets, respectively, and
$h_{j,j+1}^{(p)}(s)
=-J_{z,p}(s)Z_jZ_{j+1}
-J_{x,p}(s)X_jX_{j+1}$.
The couplings are ramped according to
$J_{\alpha,\mathrm{e}}(s)=(1-s)J_d+sJ_\alpha$
and $J_{\alpha,\mathrm{o}}(s)=sJ_\alpha$,
with $\alpha=x,z$.
We choose $J_d=J_z=1$, $J_x=0.5$,
and $s_d=(d-1/2)/8$ for $d=1,\ldots,8$,
with $\tau=T/8=0.5$ and $T=4$.
Thus, successive cycles progressively couple the initially
independent dimers while approaching uniform
ferromagnetic interactions. Further details are provided in the Supplementary Material.

Fig.~\ref{fig:vbs_dynamics}(a) shows that the mean VBS
estimator $\langle\bar O'_{\mathrm{VBS}}\rangle$ decreases
with driving depth, while the FM fluctuation amplitude
$\sqrt{\langle\bar O_{\mathrm{zFM}}^2\rangle}$ increases.
The joint distributions
$P(\bar O_{\mathrm{zFM}},\bar O'_{\mathrm{VBS}})$
in Fig.~\ref{fig:vbs_dynamics}(b) further reveal a
redistribution of weight toward weaker VBS correlations
and enhanced FM fluctuations.
These results demonstrate that the engineered drive
suppresses the initial bond dimerization while enhancing
FM fluctuations, without necessarily establishing
long-range ferromagnetic order.

{\bf Discussion.}
We have demonstrated a quantum-hardware probe of a Landau-forbidden FM--VBS transition in a frustrated one-dimensional spin chain \cite{JiangMotrunich2019PRB,Roberts2019PRB,Huang2019PRB,Mudry2019PRB,ZhangLevin2023PRL,XiYu2022CPB,Romen2024SciPostCore,Pronk2025ArXiv}.
The hardware measurements resolve the suppression of long-range ferromagnetic correlations and the redistribution of the competing FM and VBS fluctuations near the critical region.
These results should be viewed as a finite-size and finite-depth probe of competing-order criticality, with the extracted correlations affected by system size, finite bond dimension, circuit depth, sampling fluctuations, and hardware noise \cite{Preskill2018Quantum,Cerezo2021NRP,Tagliacozzo2008PRB,Pollmann2009PRL}.
Nevertheless, the observed correlation trends show that near-term quantum processors can access meaningful signatures of unconventional quantum phase transitions when guided by accurate many-body target states \cite{Ran2020PRA,Rudolph2024QST,Malz2024PRL,Smith2024PRXQuantum,MansurogluSchuch2026,Scheer2025ArXiv,ChenShen2024QuantumGeometryPRB,Shen2026TopologicalCrossoversArXiv}.
This result is consistent with recent progress in tensor-network-assisted quantum algorithms, approximate quantum compilation, MPS-to-circuit preparation, and hybrid classical--quantum simulation workflows \cite{Haghshenas2022PRX,Huang2023PRA,Robertson2023AQCTensor,AnselmeMartin2024PRA,Jaderberg2025ArXiv,Scheer2025ArXiv,shen2026simulating,Shen2026FeedbackArXiv}.
More broadly, our work provides a practical route for combining classical tensor-network methods with quantum-hardware measurements to study critical phenomena beyond the conventional Landau paradigm \cite{Senthil2004Science,Senthil2004PRB,Senthil2023Review,Cui2025CPLReview,Lee2023PRL,Bombieri2025PRL}.
Future improvements in circuit depth, error mitigation, and hardware connectivity could enable sharper finite-size scaling, more sensitive emergent-symmetry diagnostics, and clearer correlation-length growth near Landau-forbidden criticality \cite{Kim2023Nature,Chowdhury2024PRR,Zhou2024PRX,DEmidio2024PRL,Deng2024PRL,Hawashin2025NatCommun,Shu2025NatCommun,Shen2025ErrorMitigationArXiv,Li2026PracticalErrorMitigationArXiv}.

{\bf Acknowledgments.~} We acknowledge the use of IBM Quantum services for this work. The views expressed are those of the authors and do not reflect the official policy or position of IBM or the IBM Quantum team. All data and code for this work are available from the corresponding author upon reasonable request.  {This work was supported by the Singapore Ministry of Education Academic Research Fund Tier-II Grant (MOE-T2EP50224-0007).}

\bibliography{reference}
\onecolumngrid
\flushbottom
\newpage
\appendix
\setcounter{equation}{0}
\setcounter{figure}{0}
\setcounter{table}{0}
\setcounter{section}{0}
\renewcommand{\theequation}{S\arabic{equation}}
\renewcommand{\thefigure}{S\arabic{figure}}
\renewcommand{\thesection}{S\arabic{section}}
\renewcommand{\thepage}{S\arabic{page}}

\newpage
\section*{Supplementary Material}
\section{Supplementary Note 1: MPS reordering}
In this work, we introduce a correlation-guided reordering procedure for the
DMRG matrix-product state (MPS). The purpose is to represent the same physical ground state as an MPS with a different physical-site ordering. The reordered MPS is then used as the target for variational circuit optimization. The central idea is that a shallow nearest-neighbor circuit can represent the state more efficiently when strongly correlated physical sites are placed close to one another in the circuit-qubit ordering.

We start from a DMRG ground state represented as an MPS in the natural physical
order,
\begin{equation}
0,1,2,\ldots,L-1 .
\end{equation}
The corresponding state is written as
\begin{equation}
|\psi\rangle
=
\sum_{s_0,\ldots,s_{L-1}}
A_0^{s_0} A_1^{s_1}\cdots A_{L-1}^{s_{L-1}}
|s_0s_1\cdots s_{L-1}\rangle ,
\end{equation}
where $A_i^{s_i}$ is the MPS tensor on physical site $i$, and
$s_i$ denotes the local spin degree of freedom.

To determine a more favorable ordering, we first compute a
two-site correlation weight matrix from the DMRG ground
state represented as an MPS. We use the connected $ZZ$
correlator as the weight,
\begin{equation}
w_{ij}
=
\left|
\langle Z_i Z_j\rangle
-
\langle Z_i\rangle \langle Z_j\rangle
\right| .
\end{equation}
These correlations are evaluated directly through MPS
tensor contractions, without reconstructing the full
many-body wavefunction. By reusing intermediate contraction
environments, the complete correlation matrix can be
obtained with computational complexity
$O(L^2 d\chi^3)$, where $d=2$ is the local Hilbert-space
dimension and $\chi$ is the MPS bond dimension.
The additional memory required for the weight matrix
scales as $O(L^2)$. Since the target MPS has already
been obtained from DMRG, constructing this matrix
introduces only a polynomial preprocessing overhead
without additional ground-state optimization.

Given the weight matrix $w_{ij}$, we seek a permutation
\begin{equation}
\pi =
(\pi_0,\pi_1,\ldots,\pi_{L-1}),
\end{equation}
where $\pi_a$ denotes the original physical site placed at the new MPS position
$a$. The inverse permutation is denoted by
\begin{equation}
\mathrm{pos}_{\pi}(i)=a
\quad
\text{if}
\quad
\pi_a=i .
\end{equation}
A useful cost function for the ordering is
\begin{equation}
\mathcal{C}(\pi)
=
\sum_{i<j}
w_{ij}
\left|
\mathrm{pos}_{\pi}(i)
-
\mathrm{pos}_{\pi}(j)
\right| .
\end{equation}
This cost is small when strongly correlated physical sites are placed close
together in the reordered MPS. The optimized ordering is therefore approximated by
\begin{equation}
\pi_{\mathrm{opt}}
\approx
\arg\min_{\pi}
\mathcal{C}(\pi).
\end{equation}

For example, if the correlation matrix shows strong long-distance pairs
\begin{equation}
(0,L-1),\quad (1,L-2),\quad (2,L-3),\ldots,
\end{equation}
then a useful reordered sequence is
\begin{equation}
\pi_{\mathrm{edge}}
=
(0,L-1,1,L-2,2,L-3,\ldots).
\end{equation}
In this order, the originally long-range correlated pairs become nearest
neighbors in the MPS and circuit ordering. Similarly, if the state has a
dominant valence-bond pattern,
\begin{equation}
(0,1),(2,3),(4,5),\ldots ,
\end{equation}
then the natural ordering is already compatible with the dimer structure.

The reordered target state is not obtained by merely relabeling the qubits; the MPS tensors are explicitly reordered. The target state in the reordered
basis is defined by
\begin{equation}
|\psi^{(\pi)}\rangle
=
\sum_{t_0,\ldots,t_{L-1}}
\psi^{(\pi)}_{t_0t_1\cdots t_{L-1}}
|t_0t_1\cdots t_{L-1}\rangle ,
\end{equation}
with
\begin{equation}
t_a = s_{\pi_a}.
\end{equation}
Equivalently,
\begin{equation}
\psi^{(\pi)}_{t_0t_1\cdots t_{L-1}}
=
\psi_{s_0s_1\cdots s_{L-1}},
\qquad
s_{\pi_a}=t_a .
\end{equation}
This is the same physical wavefunction, but expressed in the reordered physical
basis
\begin{equation}
\pi_0,\pi_1,\ldots,\pi_{L-1}.
\end{equation}

The MPS reordering is implemented through a sequence of adjacent site swaps. Starting from the original order,
\begin{equation}
(0,1,2,\ldots,L-1),
\end{equation}
we repeatedly swap neighboring MPS sites until the desired order
$\pi$ is reached. Each adjacent swap acts locally on two neighboring MPS
tensors,
\begin{equation}
\cdots A_i^{s_i} A_{i+1}^{s_{i+1}}\cdots
\longrightarrow
\cdots \widetilde{A}_{i+1}^{s_{i+1}}
\widetilde{A}_{i}^{s_i}\cdots .
\end{equation}
Operationally, we first contract the two neighboring MPS tensors into a
two-site tensor,
\begin{equation}
\Theta^{s_i s_{i+1}}_{\alpha\gamma}
=
\sum_{\beta}
A_i^{s_i}{}_{\alpha\beta}
A_{i+1}^{s_{i+1}}{}_{\beta\gamma}.
\end{equation}
The two physical indices are then exchanged,
\begin{equation}
\Theta^{s_i s_{i+1}}_{\alpha\gamma}
\rightarrow
\Theta^{s_{i+1} s_i}_{\alpha\gamma}.
\end{equation}
The resulting tensor is reshaped into a matrix and subjected to a singular-value
decomposition,
\begin{equation}
\Theta_{\alpha s_{i+1},\, s_i \gamma}
=
\sum_{\beta'}
U_{\alpha s_{i+1},\beta'}
S_{\beta'}
V_{\beta',s_i\gamma}.
\end{equation}
The tensors are then split back into a valid MPS representation,
\begin{equation}
\widetilde{A}_{i}^{s_{i+1}}{}_{\alpha\beta'}
=
U_{\alpha s_{i+1},\beta'},
\end{equation}
and
\begin{equation}
\widetilde{A}_{i+1}^{s_i}{}_{\beta'\gamma}
=
S_{\beta'}V_{\beta',s_i\gamma}.
\end{equation}
During this procedure, small singular values may be truncated according to a
chosen cutoff $\epsilon$ or a maximum bond dimension $\chi_{\max}$. The
discarded weight at a given swap is
\begin{equation}
\epsilon_{\mathrm{disc}}
=
\sum_{\beta'>\chi_{\max}}
S_{\beta'}^2 .
\end{equation}

\section{Supplementary Note 2: Conversion of the matrix-product state into a quantum circuit}
\label{app:mps_to_circuit}

We convert the DMRG matrix-product state (MPS) into a shallow
nearest-neighbor quantum circuit using classical variational
disentanglement (CVD) \cite{MansurogluSchuch2026}. The
corresponding brick-wall matrix-product disentangling routine is denoted
BMPD. The method constructs a circuit that approximately disentangles the
target MPS into a product state. Reversing this circuit then gives a
preparation circuit for the original many-body state.

This method differs from a direct sequential embedding of the MPS tensors into unitary matrices. Its optimization target is the entanglement remaining in the MPS after the
application of local two-qubit gates, rather than the global many-body
fidelity of a variational circuit.

\subsubsection{Canonical MPS representation}

Before using the reordered DMRG state as a circuit target, we express the finite MPS in the canonical Vidal
$\Gamma$--$\Lambda$ form.
This representation separates the local tensor degrees of freedom from the entanglement spectrum on each bond, making it straightforward to quantify how much entanglement is generated across different cuts of the reordered chain.
The state is written as
\begin{equation}
|\psi_{\mathrm{MPS}}\rangle
=
\sum_{\{s_j\}}
\Gamma^{[0]s_0}
\Lambda^{[1]}
\Gamma^{[1]s_1}
\Lambda^{[2]}
\cdots
\Lambda^{[N-1]}
\Gamma^{[N-1]s_{N-1}}
|s_0s_1\cdots s_{N-1}\rangle ,
\label{eq:vidal_form_appendix}
\end{equation}
where $N$ is the number of physical sites and $s_j\in\{0,1\}$ labels the local spin state on site $j$.
The tensors $\Gamma^{[j]s_j}$ contain the local variational degrees of freedom, while the diagonal matrices $\Lambda^{[b]}$ contain the Schmidt coefficients across bond $b$.

The entries $\lambda^{[b]}_\nu$ of $\Lambda^{[b]}$ are the Schmidt coefficients for the bipartition
\begin{equation}
\{0,\ldots,b-1\}\,|\,\{b,\ldots,N-1\}.
\end{equation}
They quantify how strongly the two sides of the chain are entangled across that bond.
The corresponding $\alpha$-Rényi entanglement entropy is
\begin{equation}
S_{\alpha}^{[b]}
=
\frac{1}{1-\alpha}
\ln
\left[
\sum_{\nu}
\left(\lambda_{\nu}^{[b]}\right)^{2\alpha}
\right],
\label{eq:bond_renyi_appendix}
\end{equation}
where the Schmidt coefficients are normalized such that
\begin{equation}
\sum_{\nu}\left(\lambda_{\nu}^{[b]}\right)^2=1 .
\end{equation}
In the present calculations, we use $\alpha=2$, so that
\begin{equation}
S_{2}^{[b]}
=
-\ln
\left[
\sum_{\nu}
\left(\lambda_{\nu}^{[b]}\right)^4
\right].
\end{equation}
This bond entropy directly characterizes how difficult a reordered MPS is to approximate with a nearest-neighbor circuit: lower entanglement across the reordered cuts generally corresponds to a more circuit-compatible target state.

\subsubsection{Brick-wall disentangling circuit}

We use a brick-wall circuit made of nearest-neighbor two-qubit disentanglers.
The circuit reduces the entanglement of the reordered MPS through local two-site unitaries applied on alternating bonds.
The nearest-neighbor bonds are divided into two non-overlapping sets,
\begin{align}
\mathcal{B}_{0}
&=
\{(0,1),(2,3),(4,5),\ldots\},
\\
\mathcal{B}_{1}
&=
\{(1,2),(3,4),(5,6),\ldots\}.
\end{align}
The gates within each set act on disjoint pairs and can therefore be applied in parallel.
In our implementation, one value of \texttt{num\_layers} corresponds to one complete brick-wall round, consisting of both the even-bond and odd-bond columns.
For the default parity order $(0,1)$, the even-bond column is applied first, followed by the odd-bond column.

For round $\mu$, we define the two columns as
\begin{align}
D_{0}^{(\mu)}
&=
\prod_{i\ {\rm even}}
G_{i,i+1}^{(\mu)},
\\
D_{1}^{(\mu)}
&=
\prod_{i\ {\rm odd}}
G_{i,i+1}^{(\mu)}.
\end{align}
Since the even column is applied first, one complete brick-wall round is represented by
\begin{equation}
D^{(\mu)}
=
D_{1}^{(\mu)}D_{0}^{(\mu)} ,
\end{equation}
where the rightmost operator acts first.
After $N_{\mathrm{lay}}$ complete rounds, the full brick-wall disentangler is
\begin{equation}
D_{\mathrm{BW}}
=
D^{(N_{\mathrm{lay}})}
D^{(N_{\mathrm{lay}}-1)}
\cdots
D^{(1)} .
\label{eq:full_bw_disentangler}
\end{equation}

For an open chain of $N$ sites, each complete round acts once on every nearest-neighbor bond and therefore contains $N-1$ two-qubit gates.
Since each round consists of two parallel two-qubit columns, the two-qubit depth before decomposition into the native hardware gate set is
\begin{equation}
d_{\mathrm{2q}}
=
2N_{\mathrm{lay}},
\end{equation}
and the total number of two-qubit unitaries is
\begin{equation}
N_{\mathrm{2q}}
=
N_{\mathrm{lay}}(N-1).
\end{equation}

\subsubsection{Two-qubit gate parametrization}

Each nearest-neighbor disentangler $G_{i,i+1}$ is represented by a compact nine-parameter two-qubit unitary.
For a neighboring pair $(i,i+1)$, we write
\begin{equation}
G_{i,i+1}(\boldsymbol{\theta})
=
U_{\mathrm{ent}}
(\theta_0,\theta_1,\theta_2)
\left[
u_i(\theta_3,\theta_4,\theta_5)
\otimes
u_{i+1}(\theta_6,\theta_7,\theta_8)
\right].
\label{eq:bmpd_gate_appendix}
\end{equation}
Here, the first factor is an entangling two-qubit operation,
\begin{equation}
U_{\mathrm{ent}}
=
e^{-i\theta_0 X\otimes X}
e^{-i\theta_1 Y\otimes Y}
e^{-i\theta_2 Z\otimes Z},
\end{equation}
while the second factor applies independent single-qubit rotations to the two sites.
The local single-qubit unitary is defined as
\begin{equation}
u(v_x,v_y,v_z)
=
\exp\left[
-i\left(v_xX+v_yY+v_zZ\right)
\right].
\end{equation}
Thus, each two-qubit block contains three entangling parameters and six local rotation parameters.

\subsubsection{Local entropy minimization}

We now describe how each two-qubit disentangler is chosen.
The goal is to find a local unitary gate that reduces the entanglement across the bond on which it acts.
For a neighboring pair of sites $(i,i+1)$, we first combine the two local MPS tensors and the adjacent Schmidt spectra into a single two-site tensor,
\begin{equation}
\Theta^{[i]}_{a s t c}
=
\Lambda^{[i]}_a
\sum_b
\Gamma^{[i]s}_{a b}
\Lambda^{[i+1]}_b
\Gamma^{[i+1]t}_{b c}
\Lambda^{[i+2]}_c .
\label{eq:local_theta_appendix}
\end{equation}
Here $s$ and $t$ are the physical indices of the two sites, while $a$ and $c$ label the Schmidt states on the left and right of the two-site block.
Thus, $\Theta^{[i]}$ contains the local wavefunction information required to optimize the active bond.

\begin{figure}
 \centering
 \includegraphics[width=0.5\linewidth]{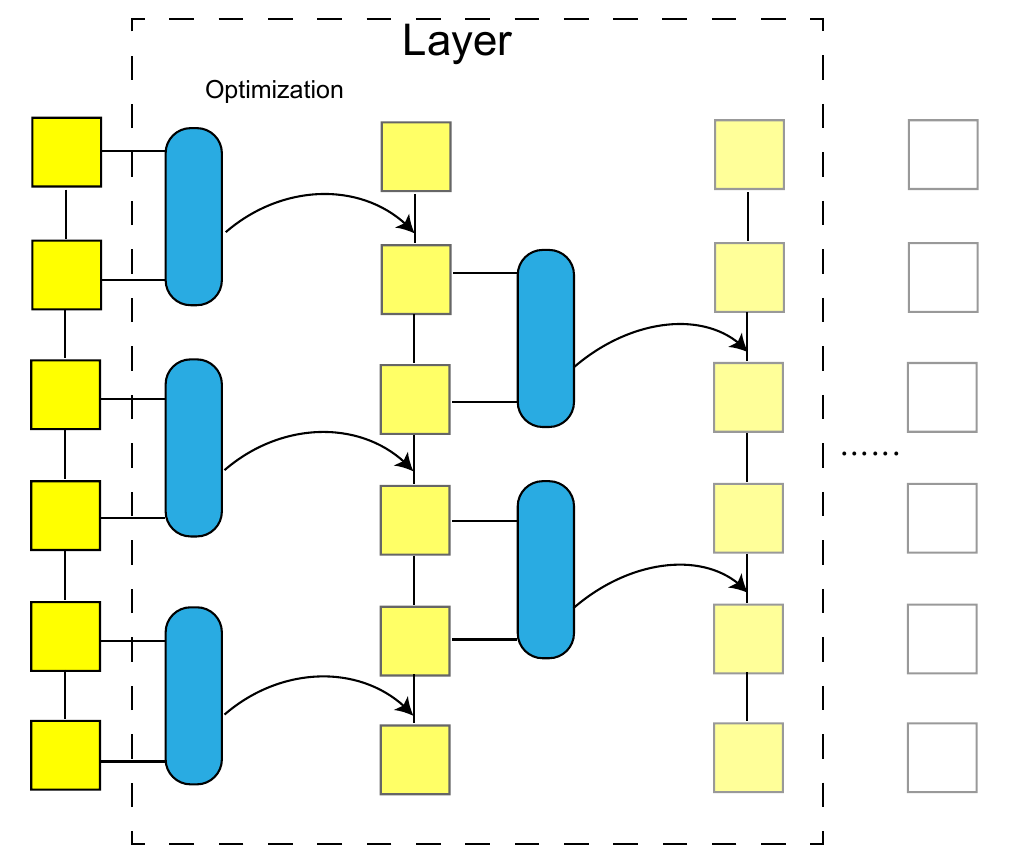}
 \caption{Structure of the BMPD optimization process. Each blue block [Eq.~\ref{eq:bmpd_gate_appendix}] is optimized to reduce the local entropy [Eq.~\ref{eq:local_cvd_cost}]. Across successive layers, the MPS becomes progressively disentangled.}
 \label{fig:circuitlayer}
\end{figure}

A trial two-qubit gate $G_{i,i+1}(\boldsymbol{\theta})$ acts only on the two physical indices,
\begin{equation}
\widetilde{\Theta}^{[i]}_{a s' t' c}
=
\sum_{s,t}
\left[
G_{i,i+1}(\boldsymbol{\theta})
\right]_{s't',st}
\Theta^{[i]}_{a s t c}.
\label{eq:gate_on_theta_appendix}
\end{equation}
To evaluate how much entanglement remains across the central bond, we reshape the transformed tensor into a matrix,
\begin{equation}
M_{(a s'),(t' c)}
=
\widetilde{\Theta}^{[i]}_{a s't'c},
\end{equation}
and perform a singular-value decomposition,
\begin{equation}
M
=
X\,\Sigma\,Y .
\label{eq:theta_svd_appendix}
\end{equation}
The singular values in $\Sigma$ give the Schmidt spectrum across the updated bond.
After normalization,
\begin{equation}
\widetilde{\lambda}_k
=
\frac{\Sigma_{kk}}
{\sqrt{\sum_l \Sigma_{ll}^{\,2}}},
\end{equation}
they define the trial Rényi entropy,
\begin{equation}
S_{\alpha}^{[i]}(\boldsymbol{\theta})
=
\frac{1}{1-\alpha}
\ln
\left[
\sum_k
\widetilde{\lambda}_k^{\,2\alpha}
\right].
\label{entropy}
\end{equation}
The optimal local gate is obtained by minimizing this entropy, as illustrated in Fig.~\ref{fig:circuitlayer}:
\begin{equation}
\boldsymbol{\theta}^{\star}_{i}
=
\operatorname*{arg\,min}_{\boldsymbol{\theta}}
S_{\alpha}^{[i]}(\boldsymbol{\theta}) .
\label{eq:local_cvd_cost}
\end{equation}
In the present calculations, we use $\alpha=2$, so this procedure directly minimizes the second Rényi entropy across the active bond.

The optimization is performed sequentially for the gates in the brick-wall circuit.
For each bond, we initialize the gate parameters from both the identity gate and a set of random initial conditions, and retain the solution that gives the smallest local Rényi entropy.
After a gate is optimized, it is immediately applied to the current MPS before moving to the next bond.
Each subsequent optimization step therefore acts on the updated state.

Finally, the optimized two-site tensor is split back into canonical MPS form.
Using the SVD in Eq.~\eqref{eq:theta_svd_appendix}, the updated Vidal tensors are
\begin{align}
\widetilde{\Gamma}^{[i]s}_{a k}
&=
\frac{X_{(a s),k}}
{\Lambda^{[i]}_a},
\\
\widetilde{\Lambda}^{[i+1]}_k
&=
\widetilde{\lambda}_k,
\\
\widetilde{\Gamma}^{[i+1]t}_{k c}
&=
\frac{Y_{k,(t c)}}
{\Lambda^{[i+2]}_c}.
\label{eq:updated_vidal_tensors}
\end{align}

\subsubsection{Termination at a product state}

After the brick-wall disentangling circuit has reduced the entanglement of the reordered MPS, the remaining state must be converted into a product-state reference.
In the original CVD construction of Ref.~\cite{MansurogluSchuch2026}, this final step is performed by projecting the residual MPS to bond dimension one through a sequential truncation map.
The resulting single-site tensors define a product state, which can then be converted into the final layer of single-qubit gates.

Our implementation follows the same goal but uses a slightly different termination procedure.
After applying the optimized brick-wall disentangler, the residual state is
\begin{equation}
|\psi_{\mathrm{res}}\rangle
=
D_{\mathrm{BW}}
|\psi_{\mathrm{MPS}}\rangle .
\label{eq:residual_mps_appendix}
\end{equation}
Rather than immediately truncating every bond to dimension one, we search for the product state
\begin{equation}
|\Phi_{\star}\rangle
=
\bigotimes_{i=0}^{N-1}|v_i\rangle
\end{equation}
that has the largest overlap with the residual MPS.
Equivalently, we maximize
\begin{equation}
F_{\mathrm{prod}}
=
\left|
\left\langle
\Phi_{\star}
\middle|
\psi_{\mathrm{res}}
\right\rangle
\right|^2 .
\end{equation}
This procedure yields the closest product-state approximation to the residual state within the chosen optimization scheme.

The maximization is carried out by alternating least squares.
During the update of site $i$, all other single-site states are held fixed, and the left and right MPS environments are contracted to obtain the optimal two-component vector $|v_i\rangle$.
The algorithm then sweeps repeatedly from left to right and right to left until convergence.
To reduce the chance of converging to a poor local optimum, we repeat this procedure for several initial product states and keep the solution with the largest value of $F_{\mathrm{prod}}$.
This product-state search is an implementation-specific refinement and should be distinguished from the direct bond-dimension-one truncation used in the original CVD formulation.

For each optimized local state,
\begin{equation}
|v_i\rangle
=
\begin{pmatrix}
a_i\\
b_i
\end{pmatrix},
\qquad
|a_i|^2+|b_i|^2=1,
\end{equation}
we construct a single-qubit unitary
\begin{equation}
R_i
=
\begin{pmatrix}
a_i^\ast & b_i^\ast\\
-b_i & a_i
\end{pmatrix}.
\label{eq:local_product_rotation}
\end{equation}
By construction, this rotation maps the local state to the computational basis state,
\begin{equation}
R_i|v_i\rangle=|0\rangle .
\end{equation}
The final product-state rotation is therefore
\begin{equation}
R_{\mathrm{f}}
=
\bigotimes_{i=0}^{N-1}R_i .
\end{equation}
Combining this rotation with the brick-wall disentangler gives the full disentangling transformation
\begin{equation}
D_{\mathrm{full}}
=
R_{\mathrm{f}}D_{\mathrm{BW}} .
\label{eq:full_disentangler_appendix}
\end{equation}
Thus, the complete transformation approximately maps the target MPS to the trivial computational product state,
\begin{equation}
D_{\mathrm{full}}
|\psi_{\mathrm{MPS}}\rangle
\simeq
|0\rangle^{\otimes N}.
\end{equation}
Equivalently, the state-preparation circuit used to approximate the target MPS is obtained by applying the inverse transformation,
\begin{equation}
|\psi_{\mathrm{MPS}}\rangle
\simeq
D_{\mathrm{full}}^{\dagger}
|0\rangle^{\otimes N}.
\end{equation}

\subsubsection{Construction and ordering of the preparation circuit}

We now describe how the disentangling circuit is converted into a state-preparation circuit.
During the disentangling procedure, the optimized two-qubit gates are stored in their order of application to the MPS.
Including the final product-state rotation, the full disentangling transformation is
\begin{equation}
D_{\mathrm{full}}
=
R_{\mathrm{f}}D_{\mathrm{BW}} .
\end{equation}
This transformation approximately maps the target MPS to the computational product state,
\begin{equation}
D_{\mathrm{full}}|\psi_{\mathrm{MPS}}\rangle
\simeq
|0\rangle^{\otimes N}.
\end{equation}

The preparation circuit is obtained by reversing this transformation.
Starting from $|0\rangle^{\otimes N}$, the desired state is prepared by applying
\begin{equation}
U_{\mathrm{prep}}
=
D_{\mathrm{full}}^\dagger
=
D_{\mathrm{BW}}^\dagger R_{\mathrm{f}}^\dagger .
\end{equation}

Thus, the single-qubit product-state rotations are placed at the beginning of the preparation circuit.
The saved two-qubit disentanglers are then traversed in reverse order, and each gate is replaced by its Hermitian conjugate.
This ordering ensures that the preparation circuit implements the inverse of the optimized disentangler and therefore maps the computational product state back to the approximate MPS target,
\begin{equation}
U_{\mathrm{prep}}|0\rangle^{\otimes N}
\simeq
|\psi_{\mathrm{MPS}}\rangle .
\end{equation}

\begin{figure}
 \centering
 \includegraphics[width=0.9\linewidth]{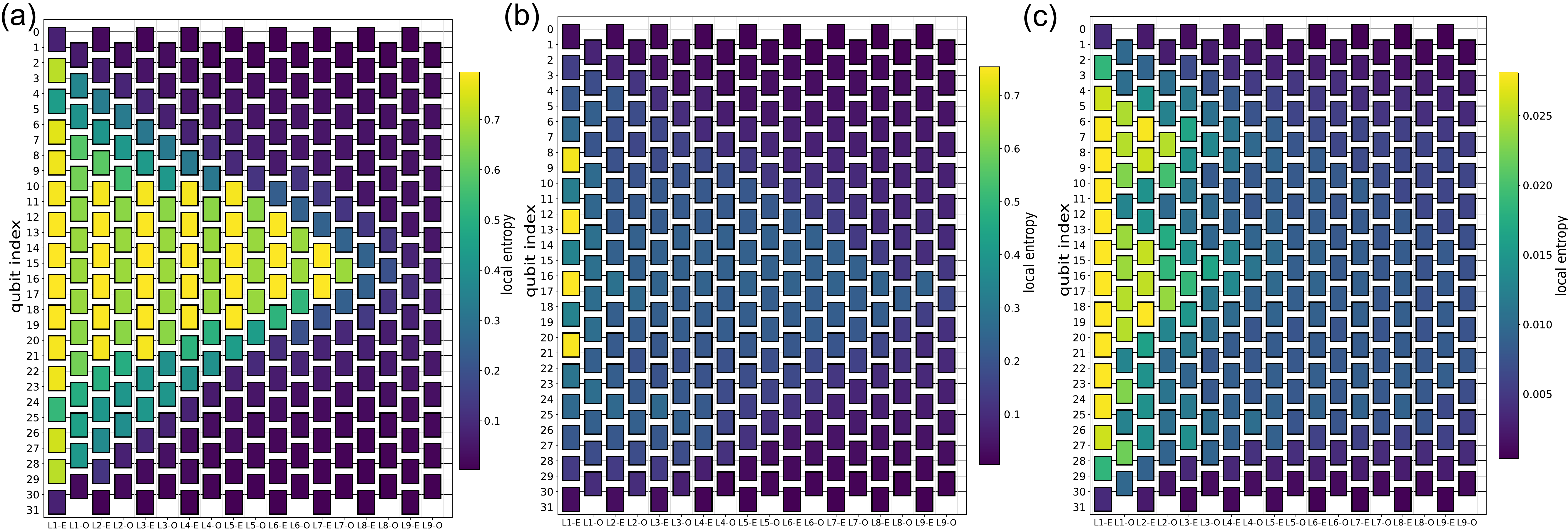}
 \caption{
Circuit-layout visualization of the local entanglement during BMPD optimization of the reordered $L=32$ MPS ($J_{z}=1.0$, $J_{x}=0.8$, $K=0.2,0.3,0.4$ for (a-c)).
Each colored block represents a two-qubit gate acting on a nearest-neighbor bond in the reordered MPS basis.
The horizontal axis labels the even (E) and odd (O) brick-wall sublayers for nine circuit layers, while the vertical axis gives the reordered qubit index.
The color encodes the local R\'enyi-2 entropy [Eq.~\ref{entropy}] after each bond update, with lower values indicating stronger local disentangling.
}
 \label{fig:local_entropy_circuit}
\end{figure}

In practice, a modest circuit depth is sufficient for the ground-state targets considered here.
For the system sizes and parameters studied here, circuits with approximately five to ten variational layers typically reach stable convergence, with optimized fidelities in the range $0.90$--$0.98$. Figure~\ref{fig:local_entropy_circuit} illustrates the spatial evolution of the local entanglement during the BMPD optimization of the reordered $L=32$ MPS.
The block color denotes the local R\'enyi-2 entropy obtained after the corresponding bond update.
At the beginning of the optimization, the larger entropy values are concentrated mainly in the interior of the chain.
As additional circuit layers are applied, the local entropies are progressively suppressed across most bonds and become more spatially uniform.

For each target state, we do not rely on a single optimized circuit.
Instead, we train a set of variational circuits using different initial parameters and retain the circuits that converge to comparable high fidelities.
We then execute this ensemble of trained circuits on the quantum processor and estimate observables by averaging over the corresponding measurement results,
\begin{equation}
\overline{O}
=
\frac{1}{N_{\mathrm{circ}}}
\sum_{\mu=1}^{N_{\mathrm{circ}}}
O_{\mu},
\label{eq:circuit_ensemble_average}
\end{equation}
where $O_{\mu}$ is the measured observable from the $\mu$th trained circuit and
$N_{\mathrm{circ}}$ is the number of accepted circuits.

This circuit-ensemble procedure provides a practical form of error averaging.
Because different trained circuits distribute the variational parameters differently, their hardware errors differ at the observable level, and averaging suppresses circuit-specific fluctuations.
Moreover, the spread among the accepted circuits provides an estimate of the systematic uncertainty associated with imperfect optimization.

\begin{figure}
    \centering
    \includegraphics[width=0.5\linewidth]{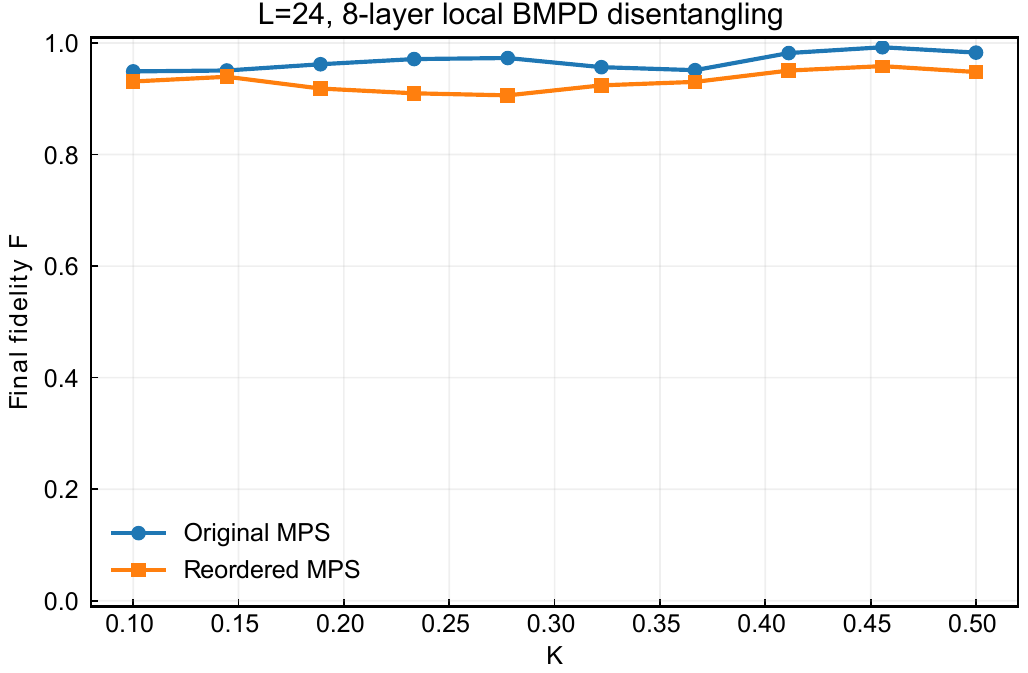}
 \caption{
\textbf{Comparison of circuit preparation fidelities
with and without MPS reordering.}
Final state-preparation fidelity $F$ as a function
of the frustration strength $K$ for the original
and reordered MPS representations at $L=24$.
Both states are prepared using eight layers of
local brick-wall matrix-product disentangling
(BMPD) optimization with the same circuit depth.
The original MPS consistently achieves slightly
higher fidelity, while the reordered MPS also
maintains high fidelity across the investigated
parameter range.
The results indicate that both representations
can be approximated accurately by shallow
nearest-neighbor circuits, although reordering
does not improve the preparation fidelity in
this comparison.
}
    \label{fig:fidelity_vs_K}
\end{figure}

\subsection*{Effect of MPS reordering on circuit preparation}

To assess the effect of site reordering on the
state-preparation performance, we compare the
circuit fidelities obtained from the original
and reordered MPS representations using the
same local BMPD optimization procedure.
Figure~\ref{fig:fidelity_vs_K}
shows the final preparation fidelity
$F=|\langle\psi_{\mathrm{target}}|
\psi_{\mathrm{circuit}}\rangle|^2$
as a function of the frustration strength $K$
for $L=24$, with eight disentangling layers
in both cases. Both representations yield high preparation
fidelities throughout the investigated range,
with $F$ approximately between $0.90$ and $1.00$.
The original MPS consistently exhibits slightly
higher fidelity than the reordered representation,
while the latter remains above approximately $0.90$.
These results demonstrate that the local BMPD
procedure can accurately approximate the target
many-body states at a fixed, relatively shallow
circuit depth.

\section{Supplementary Note 3: Finite-size scaling form of the correlator}
\label{app:czz_scaling}

We describe the fitting form used for the distance-resolved spin correlator near the FM--VBS transition.
The correlator is measured by fixing one reference spin at the center of the chain,
\begin{equation}
C_{zz}(r)
=
\left\langle
\sigma^z_{i_0}
\sigma^z_{i_0+r}
\right\rangle ,
\qquad
i_0=\frac{L}{2},
\label{eq:app_Czz_r_def}
\end{equation}
and the fit is applied to its magnitude $|C_{zz}(r)|$ over the finite-distance window shown in Fig.~\ref{fig.fig4}.

At a continuous quantum critical point, the equal-time correlator of the order parameter is expected to decay algebraically in the thermodynamic limit,
\begin{equation}
|C_{zz}(r)|
\sim
r^{-\eta_z},
\qquad
1\ll r \ll L ,
\label{eq:app_critical_power_law}
\end{equation}
where $\eta_z$ is the scaling exponent associated with the $z$-ferromagnetic order parameter.
Away from the exact critical point, the divergent length scale is replaced by a finite correlation length $\xi(K)$.
For a finite chain, the corresponding scaling form can be written as
\begin{equation}
|C_{zz}(r,K,L)|
=
r^{-\eta_z}
\mathcal{F}
\left(
\frac{r}{\xi(K)},
\frac{r}{L}
\right),
\label{eq:app_finite_size_scaling}
\end{equation}
where $\mathcal{F}$ is a nonuniversal scaling function.
In the critical limit, $\xi(K)\rightarrow\infty$ and $r/L\rightarrow0$, Eq.~\eqref{eq:app_finite_size_scaling} reduces to the algebraic form in Eq.~\eqref{eq:app_critical_power_law}.

The exponential cutoff used in the main text can be understood as the leading finite-correlation-length correction to this scaling form.
This form can be motivated by expressing the correlator as a superposition of exponential contributions,
\begin{equation}
|C_{zz}(r)|
=
\int_{\lambda_{\min}}^\infty
d\lambda\,
\rho(\lambda)
e^{-\lambda r},
\label{eq:app_spectral_representation}
\end{equation}
where $\lambda$ denotes an inverse correlation length and $\rho(\lambda)$ is the corresponding spectral weight.
At criticality, the lower edge satisfies $\lambda_{\min}=0$.
The algebraic decay in Eq.~\eqref{eq:app_critical_power_law} is reproduced if the low-$\lambda$ spectral weight has the threshold form
\begin{equation}
\rho(\lambda)
\simeq
B\lambda^{\eta_z-1},
\qquad
\lambda\rightarrow0^+ .
\label{eq:app_threshold_critical}
\end{equation}
The resulting Laplace transform gives
\begin{align}
|C_{zz}(r)|
&\simeq
B
\int_0^\infty
d\lambda\,
\lambda^{\eta_z-1}
e^{-\lambda r}
\nonumber\\
&=
B\Gamma(\eta_z)
r^{-\eta_z}.
\label{eq:app_laplace_power_law}
\end{align}

For a state with finite correlation length, the lower edge is shifted from zero to
\begin{equation}
\lambda_{\min}=\xi^{-1}.
\label{eq:app_lambda_min_finite}
\end{equation}
Keeping the same threshold exponent near this shifted edge, we write
\begin{equation}
\lambda=\xi^{-1}+\epsilon ,
\qquad
\rho(\xi^{-1}+\epsilon)
\simeq
B\epsilon^{\eta_z-1},
\qquad
\epsilon\rightarrow0^+ .
\label{eq:app_threshold_shifted}
\end{equation}
Substituting Eq.~\eqref{eq:app_threshold_shifted} into Eq.~\eqref{eq:app_spectral_representation} yields
\begin{align}
|C_{zz}(r)|
&\simeq
B
\int_0^\infty
d\epsilon\,
\epsilon^{\eta_z-1}
e^{-(\xi^{-1}+\epsilon)r}
\nonumber\\
&=
B e^{-r/\xi}
\int_0^\infty
d\epsilon\,
\epsilon^{\eta_z-1}
e^{-\epsilon r}
\nonumber\\
&=
B\Gamma(\eta_z)
r^{-\eta_z}
e^{-r/\xi}.
\label{eq:app_power_exp_derivation}
\end{align}
Thus, the leading finite-correlation-length form is
\begin{equation}
|C_{zz}(r)|
=
A r^{-\eta_z} e^{-r/\xi},
\qquad
A=B\Gamma(\eta_z).
\label{eq:app_final_fit_form}
\end{equation}
For the finite-size fits, we also include a small residual background,
\begin{equation}
|C_{zz}(r)|
=
A r^{-\eta_z} e^{-r/\xi}
+
C_0 ,
\label{eq:app_final_fit_form_with_background}
\end{equation}
where $C_0$ accounts for finite-size offsets, finite circuit depth, and residual noise in the measured correlator.

Eq.~\eqref{eq:app_final_fit_form} has the correct critical limit: as $\xi\rightarrow\infty$, the exponential factor approaches unity and the algebraic critical form is recovered.
For finite $\xi$, the same algebraic decay is cut off beyond distances of order $\xi$.
Therefore, the fitted $\eta_z$ and $\xi$ should be interpreted as effective parameters over the accessible fitting window, rather than thermodynamic-limit critical exponents.
For the IBM hardware data, the effective correlation length also includes additional suppression of long-distance correlations due to finite circuit depth, imperfect optimization, sampling uncertainty, and hardware noise.
\newpage

\section{Supplementary Note 4: IBM Quantum hardware}

\begin{figure}[h]
 \centering
 \includegraphics[width=0.7\linewidth]{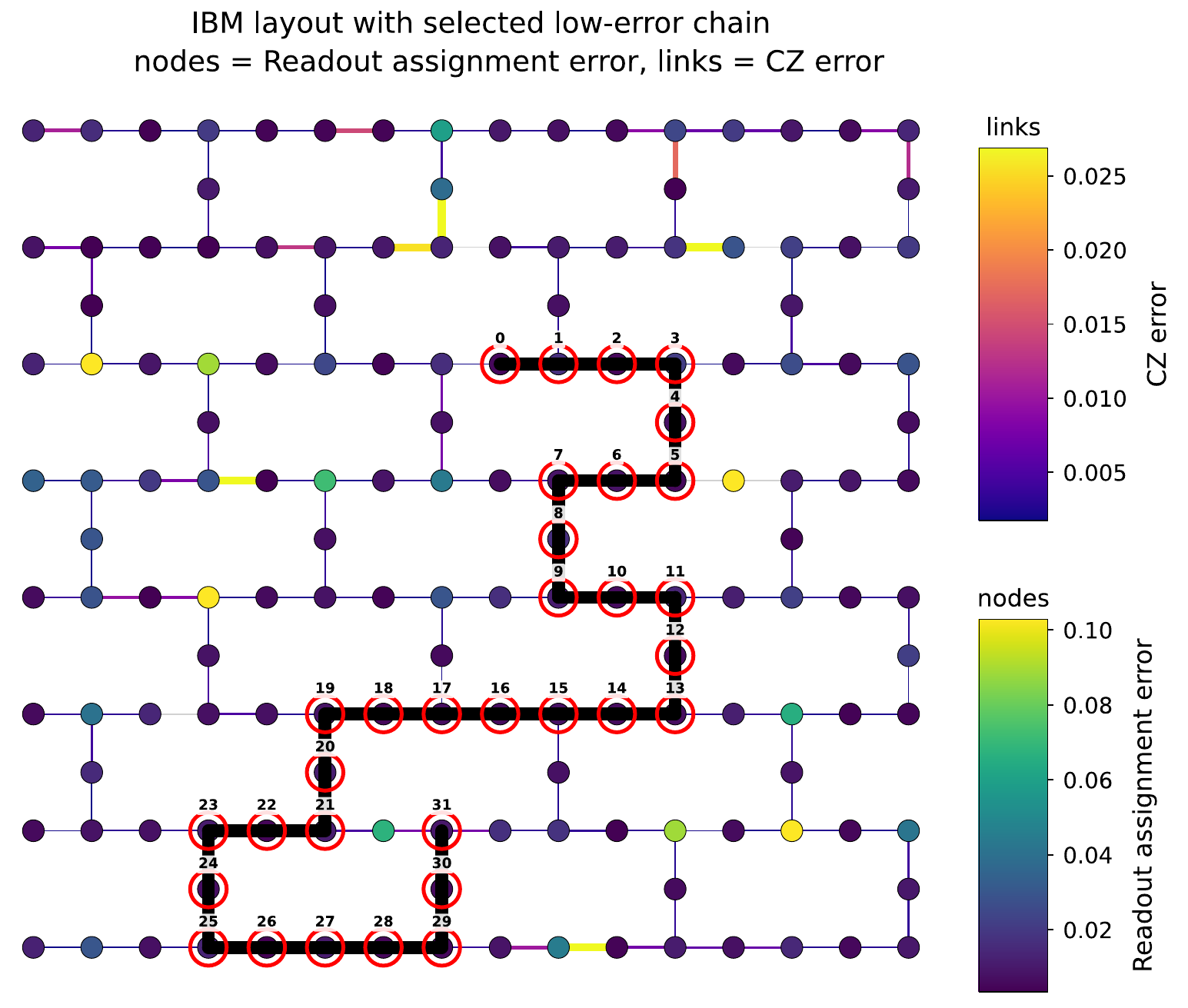}
 \caption{
\textbf{Calibration-aware selection of low-error qubit chains on IBM Quantum Fez.}
Hardware connectivity graph of the IBM Fez processor, where each node represents a physical qubit and each link represents an available two-qubit coupling.
Node colors encode the single-qubit readout assignment error, while link colors encode the corresponding CZ-gate error, as indicated by the color bars.
The selected low-error linear chains are highlighted by red open circles, and the black numbers label the ordering of qubits along each chain.
The chain is chosen by minimizing a combined calibration cost that includes both accumulated readout errors on the selected qubits and CZ errors on the links connecting neighboring qubits in the chain.
}
 \label{fig:layout_IBM}
\end{figure}

The quantum circuits are executed on the IBM Quantum Fez processor.
Since the device has a fixed hardware connectivity graph and nonuniform calibration errors across qubits and couplers, the mapping of the one-dimensional model onto physical qubits is an important part of the implementation.
We therefore use a calibration-aware layout selection procedure before submitting the circuits to hardware.

Figure~\ref{fig:layout_IBM} shows the calibration-aware chain selection on IBM Quantum Fez. For a target chain of length $L$, we search for a connected path $\mathcal{P}=(q_0,q_1,\ldots,q_{L-1})$ on the hardware coupling graph.
The quality of a candidate path is evaluated using the calibration data of the processor.
In this work, we use a cost function of the form
\begin{equation}
\mathcal{C}(\mathcal{P})
=
\sum_{i=0}^{L-1}
\epsilon_{\mathrm{ro}}(q_i)
+
\sum_{i=0}^{L-2}
\epsilon_{\mathrm{CZ}}(q_i,q_{i+1}),
\end{equation}
where $\epsilon_{\mathrm{ro}}(q_i)$ is the readout assignment error of qubit $q_i$, and
$\epsilon_{\mathrm{CZ}}(q_i,q_{i+1})$ is the calibrated CZ-gate error of the coupler connecting $q_i$ and $q_{i+1}$.
We select the connected path with the lowest calibration cost among the candidates considered.
This procedure favors physical qubits with low measurement error and nearest-neighbor couplers with low two-qubit gate error.

\section{Supplementary Note 5: Additional numerical results}

\begin{figure}
 \centering
 \includegraphics[width=0.8\linewidth]{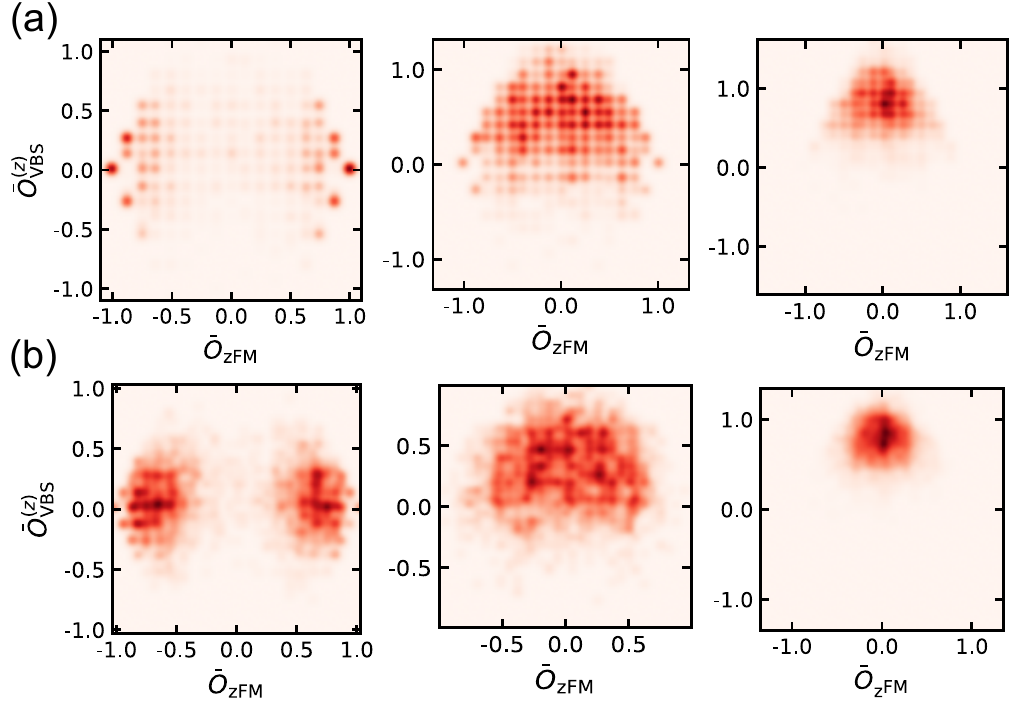}
\caption{
\textbf{Joint distributions of competing FM and VBS order parameters.}
Probability distributions of the snapshot order parameters
$(\bar O_{\rm zFM},\bar O'_{\rm VBS})$ obtained from sampled DMRG ground states across the FM--VBS transition for
\textbf{(a)} $L=32$ and \textbf{(b)} $L=64$.
From left to right, the panels correspond to the representative values of $K$ marked in Fig.~\ref{fig:fig3} of the main text and show the evolution from the FM regime through the critical region and into the VBS regime.
In the FM regime, the distribution is concentrated near finite $\bar O_{\rm zFM}$ and small $\bar O'_{\rm VBS}$, reflecting dominant ferromagnetic order.
Near the critical region, especially for the larger system size in \textbf{(b)}, the distribution broadens into an arc-like structure, indicating enhanced fluctuations between the FM and VBS directions.
In the VBS regime, the weight shifts predominantly toward the VBS axis.
}
\label{fig:joint_distribution_supp}
\end{figure}

Here we provide additional DMRG results for the joint probability distributions of the competing FM and VBS order parameters. The MPS is first expressed in the local measurement basis required by the snapshot estimators. A spin configuration $\mathbf{s}=(s_0,s_1,\ldots,s_{L-1})$ is then sampled according to the Born probability
\begin{equation}
P(\mathbf{s})
=
\left|
\left\langle
\mathbf{s}
\middle|
\psi_{\mathrm{DMRG}}
\right\rangle
\right|^2 .
\label{eq:mps_snapshot_probability}
\end{equation}
For each sampled configuration
\begin{equation}
\mathbf{s}^{(n)}
=
\left(
s_0^{(n)},s_1^{(n)},\ldots,s_{L-1}^{(n)}
\right),
\qquad
s_i^{(n)}=\pm1,
\end{equation}
where $s_i^{(n)}$ is the Pauli-$Z$ measurement outcome on site $i$, we evaluate the FM and VBS snapshot estimators directly from the resulting classical spin pattern. The ferromagnetic snapshot order parameter is the spatially averaged magnetization,
\begin{equation}
\bar O_{\mathrm{zFM}}^{(n)}
=
\frac{1}{L}
\sum_{i=0}^{L-1}
s_i^{(n)} .
\label{eq:snapshot_zfm}
\end{equation}
Thus, configurations concentrated near $\bar O_{\mathrm{zFM}}=\pm1$ correspond to the two symmetry-related $z$-polarized FM sectors, whereas configurations near $\bar O_{\mathrm{zFM}}=0$ have weak net ferromagnetic polarization.

For open boundary conditions, the local snapshot VBS estimator at the interior site $r$ is defined from the difference between the two adjacent bond products,
\begin{equation}
o_{\mathrm{VBS}}^{\prime(n)}(r)
=
(-1)^{r+r_0}
\left[
s_{r+1}^{(n)}s_r^{(n)}
-
s_r^{(n)}s_{r-1}^{(n)}
\right],
\qquad
r=1,\ldots,L-2,
\label{eq:local_snapshot_vbs}
\end{equation}
where $r_0$ fixes the convention for the staggered pattern. The corresponding spatially averaged snapshot quantity is
\begin{equation}
\bar O_{\mathrm{VBS}}^{\prime(n)}
=
\frac{1}{L-2}
\sum_{r=1}^{L-2}
o_{\mathrm{VBS}}^{\prime(n)}(r)
=
\frac{1}{L-2}
\sum_{r=1}^{L-2}
(-1)^{r+r_0}
\left[
s_{r+1}^{(n)}s_r^{(n)}
-
s_r^{(n)}s_{r-1}^{(n)}
\right].
\label{eq:snapshot_vbs}
\end{equation}
Each sampled configuration therefore produces one point
\begin{equation}
\left(
\bar O_{\mathrm{zFM}}^{(n)},
\bar O_{\mathrm{VBS}}^{\prime(n)}
\right)
\end{equation}
in the joint-order-parameter plane. Repeating the procedure for all $N_{\mathrm{samp}}$ configurations yields the joint probability distribution of the competing FM and VBS snapshot orders.

Figure~\ref{fig:joint_distribution_supp} shows the evolution of the joint distribution across the transition for $L=32$ and $L=64$.
In the FM regime, the distribution is concentrated near finite values of $\bar O_{\rm zFM}$, indicating dominant ferromagnetic ordering.
As the system approaches the critical region, the distribution spreads between the FM and VBS directions.
This broadening is more clearly visible for the larger system size, where the weight forms an arc-like structure in the
$(\bar O_{\rm zFM},\bar O'_{\rm VBS})$ plane.
This behavior is consistent with enhanced fluctuations between the competing orders near the transition.

For a coupling close to, but not exactly at, the critical point, the relevant
scaling variable is
\begin{equation}
x=(K-K_c)L^{1/\nu}.
\end{equation}
When the accessible sizes satisfy
\begin{equation}
|K-K_c|L^{1/\nu}\lesssim 1,
\end{equation}
the system remains inside the finite-size critical window and
$\mathcal{X}(x)$ varies only weakly over the studied range.  One therefore
expects
\begin{equation}
\xi_{\mathrm{2nd}}(L,K)
\simeq
c(K)L
\end{equation}
over the accessible system sizes, even when $K$ is not exactly equal to
$K_c$.  At $K=0.35$, both the DMRG and IBM hardware results display this
approximately linear increase over the system sizes considered here,
indicating that the correlation scale remains controlled primarily by the
finite system size within this window.  We therefore use the observed
$\xi_{\mathrm{2nd}}\sim L$ behavior as a finite-size signature of
critical-like correlations rather than as an independent high-precision
determination of $K_c$ or $\nu$.
\section{Supplementary Note 6: Estimation of noise effects}

For the hardware implementation, each variational layer contains two nearest-neighbor two-qubit sublayers acting on the even and odd bonds.
Each bond unitary is compiled into the native CZ basis using three CZ gates, together with single-qubit rotations. Thus each even or odd entangling sublayer corresponds to
three effective CZ layers.
For a circuit with $N_{\mathrm{lay}}$ variational layers, the effective CZ depth is therefore $D_{\mathrm{CZ}}^{\mathrm{eff}}\simeq6N_{\mathrm{lay}}$.

To estimate the accumulated two-qubit noise associated with this depth, we use
a representative CZ error rate of
\begin{equation}
p_{\mathrm{CZ}}=0.003.
\end{equation}
Within a simplified independent-layer approximation, the effective survival
factor after $D_{\mathrm{CZ}}^{\mathrm{eff}}$ sequential CZ layers is
\begin{equation}
P_{\mathrm{CZ}}^{\mathrm{eff}}
=
\left(1-p_{\mathrm{CZ}}\right)^{D_{\mathrm{CZ}}^{\mathrm{eff}}},
\label{eq:effective_cz_survival}
\end{equation}
and the corresponding accumulated effective error is
\begin{equation}
\epsilon_{\mathrm{CZ}}^{\mathrm{eff}}
=
1-
\left(1-p_{\mathrm{CZ}}\right)^{D_{\mathrm{CZ}}^{\mathrm{eff}}}.
\label{eq:effective_cz_error}
\end{equation}

For $L=32$, the circuit contains $N_{\mathrm{lay}}=10$ variational layers, giving $D_{\mathrm{CZ}}^{\mathrm{eff}}=6\times10=60$.
The corresponding depth-based survival estimate is
\begin{equation}
P_{\mathrm{CZ}}^{\mathrm{eff}}
=
0.997^{60}
\simeq
0.835.
\end{equation}

For the $L=64$ circuit, we use $N_{\mathrm{lay}}=18$, corresponding to an effective CZ depth of $D_{\mathrm{CZ}}^{\mathrm{eff}}=6N_{\mathrm{lay}}=108$.
For a representative CZ error rate of $p_{\mathrm{CZ}}=0.003$, the depth-based survival factor is
\begin{equation}
P_{\mathrm{CZ}}^{\mathrm{eff}}
=
\left(1-p_{\mathrm{CZ}}\right)^{D_{\mathrm{CZ}}^{\mathrm{eff}}}
=
0.997^{108}
\simeq
0.723.
\end{equation}
By contrast, improving the CZ error rate to $p_{\mathrm{CZ}}=0.001$ increases the estimated survival factor to
\begin{equation}
P_{\mathrm{CZ}}^{\mathrm{eff}}
=
0.999^{108}
\simeq
0.898.
\end{equation}
Thus, reducing the CZ error rate from $0.003$ to $0.001$ increases the estimated depth-based survival factor from approximately $72.3\%$ to $89.8\%$. This comparison highlights the strong sensitivity of deep circuits to even modest improvements in two-qubit gate fidelity.

\section{Supplementary Note 7: Finite-size scaling of the second-moment correlation length}
\label{sec:si_xi2_scaling}

In the main text, we characterize the long-range ferromagnetic correlations
using the following second-moment correlation length
$\xi_{\mathrm{2nd}}$ extracted from the static structure factor.
Here we provide a detailed justification for the approximately linear
finite-size dependence
\begin{equation}
\xi_{\mathrm{2nd}}(L)\propto L
\label{eq:si_xi_linear_statement}
\end{equation}
observed near the critical region.

For a chain of length $L$, we define the static ferromagnetic structure factor
as
\begin{equation}
S_{zz}(q,L,K)
=
\frac{1}{L}
\sum_{i,j=0}^{L-1}
e^{iq(i-j)}
\left\langle Z_i Z_j \right\rangle ,
\label{eq:si_szz_definition}
\end{equation}
where $K$ denotes the tuning parameter of the Hamiltonian.
The smallest nonzero momentum compatible with the finite system is $q_{\min}=\frac{2\pi}{L}$.
The second-moment correlation length is then evaluated from
\begin{equation}
\xi_{\mathrm{2nd}}(L,K)
=
\frac{1}{2\sin(\pi/L)}
\left[
\frac{S_{zz}(0,L,K)}
     {S_{zz}(q_{\min},L,K)}
-1
\right]^{1/2}.
\label{eq:si_xi2_definition}
\end{equation}
The factor $2\sin(q_{\min}/2)=2\sin(\pi/L)$ is the lattice momentum
corresponding to the continuum wave vector $q_{\min}$.
For large $L$,
\begin{equation}
2\sin\left(\frac{\pi}{L}\right)
=
\frac{2\pi}{L}
+
\mathcal{O}(L^{-3}),
\label{eq:si_lattice_momentum_expansion}
\end{equation}
so that Eq.~\eqref{eq:si_xi2_definition} reduces to the conventional
second-moment estimator in the thermodynamic limit.

Close to a continuous quantum critical point, the thermodynamic correlation
length diverges according to
\begin{equation}
\xi_{\infty}(K)
\sim
|K-K_c|^{-\nu},
\label{eq:si_bulk_xi}
\end{equation}
where $K_c$ denotes the critical coupling and $\nu$ is the correlation-length
critical exponent
\cite{Cardy1996Scaling,Sachdev2011QPT}.

For a finite system, this divergence cannot continue indefinitely because the
largest available infrared length scale is the system size $L$.
The appropriate finite-size scaling form is therefore
\begin{equation}
\xi_{\mathrm{2nd}}(L,K)
=
L\,
\mathcal{X}
\left[
(K-K_c)L^{1/\nu}
\right],
\label{eq:si_xi_fss_general}
\end{equation}
where $\mathcal{X}$ is a dimensionless scaling function.

At the critical point, $K=K_c$, 
the scaling variable vanishes and Eq.~\eqref{eq:si_xi_fss_general} becomes
\begin{equation}
\xi_{\mathrm{2nd}}(L,K_c)
=
L\,\mathcal{X}(0).
\label{eq:si_xi_at_kc}
\end{equation}
Consequently, we have
\begin{equation}
\frac{\xi_{\mathrm{2nd}}(L,K_c)}{L}
=
\mathcal{X}(0)
\label{eq:si_xi_over_L_constant}
\end{equation}
which is independent of $L$ to leading order, and hence
\begin{equation}
\xi_{\mathrm{2nd}}(L,K_c)
\propto L .
\label{eq:si_xi_linear}
\end{equation}

\subsubsection{Derivation directly from the structure factor}

Near criticality, the equal-time correlation function obeys the finite-size
scaling form
\begin{equation}
C_{zz}(r,L)
=
\langle Z_j Z_{j+r}\rangle
=
r^{-2\Delta_z}
f\left(\frac{r}{L}\right),
\label{eq:si_czz_scaling}
\end{equation}
where $\Delta_z$ is the scaling dimension of the ferromagnetic order
parameter and $f(r/L)$ is a dimensionless finite-size scaling function.
Equation~\eqref{eq:si_czz_scaling} expresses the fact that, at a critical
point, there is no intrinsic correlation length in the thermodynamic limit.
The only long-distance length scale in a finite system is therefore the
system size $L$.

An equivalent representation of Eq.~\eqref{eq:si_czz_scaling} is
\begin{equation}
C_{zz}(r,L)
=
L^{-2\Delta_z}
\mathcal{F}\left(\frac{r}{L}\right),
\label{eq:si_czz_scaling_L}
\end{equation}
where
\begin{equation}
\mathcal{F}(x)
=
x^{-2\Delta_z}f(x).
\end{equation}
This form is particularly convenient for deriving the finite-size scaling
of the structure factor. The static structure
factor is
\begin{equation}
S_{zz}(q,L)
=
\frac{1}{L}
\sum_{i,j=0}^{L-1}
e^{iq(i-j)}
C_{zz}(i-j,L).
\label{eq:si_structure_factor_double_sum}
\end{equation}
Introducing the relative coordinate $r=i-j$, the sum over the reference
site produces a factor of $L$, which cancels the normalization factor
$1/L$.  Up to the periodic identification of $r$, one therefore obtains
\begin{equation}
S_{zz}(q,L)
=
\sum_r
e^{iqr}
C_{zz}(r,L).
\label{eq:si_structure_factor_relative}
\end{equation}
Substituting Eq.~\eqref{eq:si_czz_scaling_L} gives
\begin{equation}
S_{zz}(q,L)
=
L^{-2\Delta_z}
\sum_r
e^{iqr}
\mathcal{F}\left(\frac{r}{L}\right).
\label{eq:si_structure_factor_scaling_sum}
\end{equation}

For sufficiently large $L$, the long-distance part of the discrete sum can
be written in terms of the dimensionless coordinate
\begin{equation}
x=\frac{r}{L},
\qquad
r=Lx,
\end{equation}
for which one lattice increment corresponds to $\Delta x=\frac{1}{L}.$ Consequently, $\sum_r
\longrightarrow
L\int dx,$
and Eq.~\eqref{eq:si_structure_factor_scaling_sum} becomes
\begin{align}
S_{zz}(q,L)
&\simeq
L^{1-2\Delta_z}
\int dx\,
e^{i(qL)x}
\mathcal{F}(x)
\nonumber\\
&=
L^{1-2\Delta_z}
\Phi(qL),
\label{eq:si_structure_factor_derivation}
\end{align}
where we have defined the dimensionless momentum-space scaling function $\Phi(y)
=
\int dx\,
e^{iyx}
\mathcal{F}(x).$
Thus, in one spatial dimension, the critical contribution to the structure
factor has the finite-size scaling form
\begin{equation}
S_{zz}(q,L)
=
L^{1-2\Delta_z}\Phi(qL)
\label{eq:si_szz_scaling}
\end{equation}
or, equivalently,
\begin{equation}
S_{zz}(q,L)
=
L^{y_S}\Phi(qL),
\qquad
y_S=1-2\Delta_z.
\label{eq:si_szz_scaling_exponent}
\end{equation}
More generally, in $d$ spatial dimensions the corresponding exponent is
$y_S=d-2\Delta_z$.

For the two momenta $q=0$ and
$q_{\min}=\frac{2\pi}{L}$, 
we obtain
\begin{align}
S_{zz}(0,L)
&=
L^{y_S}\Phi(0),
\label{eq:si_szero_scaling}
\\
S_{zz}(q_{\min},L)
&=
L^{y_S}\Phi(2\pi).
\label{eq:si_sqmin_scaling}
\end{align}
Therefore, the common power-law dependence on $L$ cancels in their ratio,
giving
\begin{equation}
\frac{S_{zz}(0,L)}
     {S_{zz}(q_{\min},L)}
\longrightarrow
R^\star
\equiv
\frac{\Phi(0)}{\Phi(2\pi)}.
\label{eq:si_ratio_fixedpoint}
\end{equation}
Thus, at criticality the dimensionless structure-factor ratio approaches an
$L$-independent value $R^\star$.

Substituting Eq.~\eqref{eq:si_ratio_fixedpoint} into
Eq.~\eqref{eq:si_xi2_definition} yields
\begin{equation}
\xi_{\mathrm{2nd}}(L)
=
\frac{\sqrt{R^\star-1}}
     {2\sin(\pi/L)}.
\label{eq:si_xi_ratio}
\end{equation}
Using the large-$L$ expansion
\begin{equation}
2\sin\left(\frac{\pi}{L}\right)
=
\frac{2\pi}{L}
+
\mathcal{O}(L^{-3}),
\end{equation}
we obtain
\begin{equation}
\xi_{\mathrm{2nd}}(L)
=
\frac{\sqrt{R^\star-1}}{2\pi}\,L
+
\mathcal{O}(L^{-1}).
\label{eq:si_xi_linear_structure}
\end{equation}
Hence, the linear scaling of $\xi_{\mathrm{2nd}}$ is a direct consequence of
the scale invariance of the low-momentum structure-factor ratio.

\subsection{Error mitigation for the second-moment estimator}
\label{sec:si_xi2_zne}

The second-moment correlation length depends nonlinearly on the ratio of two
low-momentum structure factors,
\begin{equation}
R_\xi
=
\frac{S_{zz}(0)}
     {S_{zz}(q_{\min})},
\qquad
q_{\min}=\frac{2\pi}{L}.
\label{eq:si_Rxi}
\end{equation}

A natural first mitigation strategy is symmetry-sector postselection, in which
measurement outcomes outside the target symmetry sector are discarded.
For the present observable, however, such postselection is not sufficient.
It can remove only errors that cause leakage between different symmetry
sectors, while hardware errors that preserve the symmetry remain entirely
inside the retained data.

This distinction is particularly important for $R_\xi$.  Suppose, for
example, that noise produces a common attenuation of the correlations,
\begin{equation}
C_{ij}^{\rm hw}
\simeq
\alpha C_{ij}^{\rm ideal}.
\end{equation}
Then
\begin{equation}
S_{zz}^{\rm hw}(q)
\simeq
\alpha S_{zz}^{\rm ideal}(q),
\end{equation}
so that the common factor cancels,
\begin{equation}
\frac{S_{zz}^{\rm hw}(0)}
     {S_{zz}^{\rm hw}(q_{\min})}
\simeq
\frac{S_{zz}^{\rm ideal}(0)}
     {S_{zz}^{\rm ideal}(q_{\min})}.
\end{equation}
Symmetry postselection therefore cannot further correct such a common-mode
error.  More importantly, the residual bias in $R_\xi$ is generated by
errors that modify the two momentum components differently,
\begin{equation}
S_{zz}^{\rm hw}(0)
\rightarrow
S_{zz}^{\rm ideal}(0)+\delta S_0,
\qquad
S_{zz}^{\rm hw}(q_{\min})
\rightarrow
S_{zz}^{\rm ideal}(q_{\min})+\delta S_q,
\end{equation}
with $\delta S_0/S_{zz}(0)\neq
\delta S_q/S_{zz}(q_{\min})$.
Such distortions can arise from symmetry-preserving gate errors and
decoherence and are not removed by selecting a symmetry sector.
Consequently, symmetry selection produces little improvement for the
second-moment estimator when the dominant hardware error remains within the
physical symmetry sector. We therefore mitigate the error directly using zero-noise
extrapolation (ZNE). We also apply ZNE to the other
quantum-hardware measurements presented in this work
to suppress the effects of hardware noise and improve
the agreement with the ideal numerical results.

\section{Supplementary Note 8: Floquet-type dynamics of competing FM and VBS correlations}
\label{sec:sn8_floquet}

Here, we detail the driven-circuit protocol used to study the evolution of competing ferromagnetic (FM) and valence-bond-solid (VBS) correlations in Fig.~\ref{fig:vbs_dynamics}. Unlike the ground-state calculations for the frustrated $XZ$ chain, this setup starts from a directly preparable dimerized state and applies a prescribed sequence of nearest-neighbor unitary gates. 

\subsection*{Initial Bell-dimer state}

We consider an open chain of even length $L$ and initialize the product of Bell pairs
\begin{equation}
|\psi_0\rangle=\bigotimes_{m=0}^{L/2-1}|\Phi^+\rangle_{2m,2m+1},
\qquad |\Phi^+\rangle=\frac{|00\rangle+|11\rangle}{\sqrt{2}}.
\label{eq:sn8_initial}
\end{equation}
Each pair can be prepared from $|00\rangle$ by applying a Hadamard gate to qubit $2m$ followed by a controlled-NOT gate from $2m$ to $2m+1$. All $L/2$ pairs can be prepared in parallel, requiring a single entangling layer when the corresponding physical couplers are available. The two spins in each dimer satisfy
$\langle X_{2m}X_{2m+1}\rangle=\langle Z_{2m}Z_{2m+1}\rangle=1$, whereas correlations between distinct dimers vanish; in particular, $\langle Z_{2m+1}Z_{2m+2}\rangle=0$. The initial bond correlations are therefore staggered. This Bell-dimer product is an \emph{idealized dimerized reference state}, not a product of singlets and not the exact VBS ground state of the frustrated Hamiltonian studied in the main text.

The state in Eq.~\eqref{eq:sn8_initial} is the unique ground state of the decoupled-dimer Hamiltonian
\begin{equation}
H_{\mathrm{dim}}=-J_d\sum_{m=0}^{L/2-1}
\bigl(X_{2m}X_{2m+1}+Z_{2m}Z_{2m+1}\bigr),\qquad J_d>0.
\label{eq:sn8_Hdim}
\end{equation}
Indeed, $|\Phi^+\rangle$ is a simultaneous $+1$ eigenstate of $XX$ and $ZZ$ on each dimer. At the other end of the driving protocol, the target interaction structure is the uniform nearest-neighbor easy-axis ferromagnet
\begin{equation}
H_{\mathrm{FM}}=-\sum_{j=0}^{L-2}
\left(J_z Z_jZ_{j+1}+J_x X_jX_{j+1}\right).
\label{eq:sn8_Hfm}
\end{equation}

\subsection*{Layer-dependent first-order driving circuit}

We partition the nearest-neighbor bonds into two nonoverlapping sets,
$\mathcal B_{\mathrm e}=\{(0,1),(2,3),\ldots,(L-2,L-1)\}$ and
$\mathcal B_{\mathrm o}=\{(1,2),(3,4),\ldots,(L-3,L-2)\}$.
The formal interpolation $H(s)=(1-s)H_{\mathrm{dim}}+sH_{\mathrm{FM}}$ gives the even- and odd-bond generators
\begin{align}
H_{\mathrm e}(s)&=-\sum_{(j,j+1)\in\mathcal B_{\mathrm e}}
\left[J_{z,\mathrm e}(s)Z_jZ_{j+1}+J_{x,\mathrm e}(s)X_jX_{j+1}\right],\nonumber\\
H_{\mathrm o}(s)&=-\sum_{(j,j+1)\in\mathcal B_{\mathrm o}}
\left[J_{z,\mathrm o}(s)Z_jZ_{j+1}+J_{x,\mathrm o}(s)X_jX_{j+1}\right],
\label{eq:sn8_HeHo}
\end{align}
with layer-dependent positive coupling magnitudes
\begin{align}
J_{\alpha,\mathrm e}(s)&=(1-s)J_d+sJ_\alpha,\nonumber\\
J_{\alpha,\mathrm o}(s)&=sJ_\alpha,\qquad \alpha\in\{x,z\}.
\label{eq:sn8_couplings}
\end{align}
Thus, the initially strong even-bond dimers persist while the odd-bond interactions progressively connect previously independent pairs. For the parameters $J_d=J_z=1$ and $J_x=0.5$, these couplings reduce to
$J_{z,\mathrm e}=1$, $J_{x,\mathrm e}=1-s/2$,
$J_{z,\mathrm o}=s$, and $J_{x,\mathrm o}=s/2$. Here, we use $N=8$ driving cycles and a nominal evolution parameter $T=4$, with a dimensionless pulse parameter $\tau=T/N=0.5$ in each bond sublayer. The coupling schedule is evaluated at the midpoint of each cycle,
\begin{equation}
s_d=\frac{d-1/2}{N},\qquad d=1,\ldots,N.
\label{eq:sn8_schedule}
\end{equation}

The two-stage unitary defining one cycle is
\begin{equation}
U_F(s_d)=U_{\mathrm o}(s_d)U_{\mathrm e}(s_d),\qquad
U_p(s)=e^{-i\tau H_p(s)},\quad p\in\{\mathrm e,\mathrm o\}.
\label{eq:sn8_floquet_cycle}
\end{equation}
The resulting depth-$d$ state is
\begin{equation}
|\psi_d\rangle=U_F(s_d)U_F(s_{d-1})\cdots U_F(s_1)|\psi_0\rangle,
\qquad |\psi_{d=0}\rangle=|\psi_0\rangle.
\label{eq:sn8_depth_state}
\end{equation}

\subsection*{Two-qubit gate implementation and circuit cost}

For either bond type, write
$h_{j,j+1}(s)=-a_z(s)Z_jZ_{j+1}-a_x(s)X_jX_{j+1}$,
with $(a_z,a_x)$ taken from Eq.~\eqref{eq:sn8_couplings}. Since $[X\otimes X,Z\otimes Z]=0$ on the same bond, its unitary factors \emph{exactly} as
\begin{equation}
\begin{split}
U_{j,j+1}(s)&=e^{-i\tau h_{j,j+1}(s)}\\
&=e^{i\tau a_x(s)X_jX_{j+1}}
  e^{i\tau a_z(s)Z_jZ_{j+1}}\\
&=R_{XX}\bigl(-2\tau a_x(s)\bigr)
  R_{ZZ}\bigl(-2\tau a_z(s)\bigr).
\end{split}
\label{eq:sn8_twoqubit}
\end{equation}
Here $R_{PP}(\theta)=e^{-i\theta P\otimes P/2}$. Accordingly, the circuit applies $R_{XX}(-2\tau a_x)$ and $R_{ZZ}(-2\tau a_z)$ to each active pair. Splitting $XX$ and $ZZ$ on the \emph{same} bond introduces no approximation; the even--odd operator ordering is the only first-order splitting of overlapping interactions.

There is one exact gate-saving simplification. On every initial Bell pair, both $XX$ and $ZZ$ have eigenvalue $+1$, so the first even-bond sublayer acts on $|\psi_0\rangle$ only through the global phase
\begin{equation}
U_{\mathrm e}(s_1)|\psi_0\rangle=
\exp\!\left[i\tau\frac L2
\bigl(J_{x,\mathrm e}(s_1)+J_{z,\mathrm e}(s_1)\bigr)\right]
|\psi_0\rangle.
\label{eq:sn8_redundant}
\end{equation}
We therefore omit this first even-bond sublayer in the submitted circuit without altering any state expectation value or measurement probability. For a depth-$d\geq1$ circuit, the number of active interaction-bond blocks before hardware decomposition is $d(L-1)-L/2$, in addition to the $L/2$ entangling gates used to prepare the initial Bell dimers.

For each computational-basis shot, let $\boldsymbol{s}=(s_0,\ldots,s_{L-1})$ with $s_j=+1$ for a measured bit $0$ and $s_j=-1$ for a measured bit $1$. As in the main text, we use the snapshot estimators
\begin{align}
\bar O_{\mathrm{zFM}}(\boldsymbol{s})
&=\frac{1}{L}\sum_{j=0}^{L-1}s_j,\nonumber\\
\bar O'_{\mathrm{VBS}}(\boldsymbol{s})
&=\frac{1}{L-2}\sum_{r=1}^{L-2}(-1)^r
\bigl(s_{r+1}s_r-s_rs_{r-1}\bigr).
\label{eq:sn8_snapshots}
\end{align}
The VBS estimator is a $Z$-snapshot-compatible staggered-bond diagnostic. It must be distinguished from the full energy-bond VBS order parameter involving both $XX$ and $ZZ$ correlators. With the sign convention in Eq.~\eqref{eq:sn8_snapshots}, the initial Bell-dimer state has $\langle\bar O'_{\mathrm{VBS}}\rangle_{d=0}=1$.  The quantities shown in Fig.~\ref{fig:vbs_dynamics}(a) are the ensemble-averaged $\langle\bar O'_{\mathrm{VBS}}\rangle_d$ and the root-mean-square FM fluctuation amplitude,
\begin{equation}
M_{\mathrm{FM}}^{\mathrm{rms}}(d)
\equiv\sqrt{\bigl\langle\bar O_{\mathrm{zFM}}^2\bigr\rangle_d},
\quad
\bigl\langle\bar O_{\mathrm{zFM}}^2\bigr\rangle_d
=\frac{1}{L^2}\left[L+2\sum_{i<j}\langle Z_iZ_j\rangle_d\right].
\label{eq:sn8_Mrms}
\end{equation}

\flushbottom
\end{document}